\documentstyle[12pt,epsf]{article}
\newcommand{\nb}{{\bf  b}}     
\newcommand{\nne}{{\bf e}}   

\newcommand{\nk}{{\bf  k}}
\newcommand{\nl}{{\bf  l}}
\newcommand{\np}{{\bf  p}}       
\newcommand{\nq}{{\bf  q}}
\newcommand{\nr}{{\bf  r}}         
\newcommand{\ns}{{\bf  s}}

\newcommand{\nw}{{\bf  w}}

\newcommand{\nM}{{\bf  M}}

\newcommand{\nS}{{\bf  S}}

\newcommand{\hp}{{\bf \hat{p}}}
 
\newcommand{\nsigma}{\mbox{\boldmath$\sigma$}}
\newcommand{\nOmega}{\mbox{\boldmath$\Omega$}}

\begin{document}

\begin{titlepage}
\mbox{} 
\vspace*{2.5\fill} 

{\Large\bf 
\begin{center}
%
Electron helicity-dependence  \\
in $(e,e'p)$ reactions with polarized nuclei\\
and the fifth response function
%
\end{center}
} 

\vspace{1\fill} 

\begin{center}
{\large J.E. Amaro$^{1}$ and T.W. Donnelly$^2$} 
\end{center}

{\small 
\begin{center}
$^1$ {\em Departamento de F\'{\i}sica Moderna, 
          Universidad de Granada,
          Granada 18071, Spain}  \\   
$^2$ {\em Center for  Theoretical  Physics,
          Laboratory   for Nuclear Science  and  
          Dept. of Physics, }\\   
     {\em Massachusetts Institute  of  Technology, 
          Cambridge,  MA  02139, U.S.A.}\\[2mm] 
\end{center}
} 

\kern 1.5 cm 

\hrule \kern 3mm 

{\small \noindent
{\bf Abstract} 
\vspace{3mm} 

The cross section for electron-induced proton knockout reactions with
polarized beam and target is computed in a DWIA model. The analysis is
focused on the electron helicity-dependent (HD) piece of the cross
section.  For the case of in-plane emission a new symmetry is found in
the HD cross section, which takes opposite values for specific
opposing pairs of nuclear orientations, explaining why the fifth
response (involving polarized electrons, but unpolarized nuclei) is
zero for this choice of kinematics.  Even when this symmetry breaks
down in the case of nucleon emission out of the scattering plane, it
is shown that it is still present in the separate response functions.
A physical explanation of these results is presented using a
semi-classical model of the reaction based on the PWIA and on the
concept of a nucleon orbit with its implied expectation values of
position and spin.  }

\kern 3mm \hrule 

\noindent
{\em PACS:} 25.30.Fj; 24.10.Eq; 24.70.+s; 29.27.Hj

\noindent
{\em Keywords:} electromagnetic nucleon knockout; polarized beam and
target; final state interactions; structure response functions

\vfill

\noindent MIT/CTP\#3188 \hfill September 2001

\end{titlepage}


\section{Introduction}


In this work we continue our systematic investigation of spin
observables in $\vec{A}(\vec{e},e'p)B$ reactions initiated in
\cite{Ama98a,Ama99a}. Relying on the general formalism for exclusive
reactions \cite{Ras89} we particularized in \cite{Ama98a} to nucleon
knock-out, performed an explicit multipole expansion of the response
functions involved and developed a DWIA model of the reaction,
including relativistic corrections \cite{Ama96b}. This was applied in
an analysis of the full set of multipoles or reduced response
functions (defined to be independent of the polarization angles) that
can be extracted in studies of such reactions.  The emphasis there was
placed on studies of the sensitivity of the predictions to variations
in the final-state interactions (FSI) and to inclusion of relativistic
contributions to the electromagnetic current.  The results found
showed that these have very different impacts on the various response
functions, which implies in the end that the importance of the
specifics of the reaction mechanisms underlying the cross section
strongly depends on the choice of polarization angles.  The
possibilities opened by this sensitivity to interesting physical
effects underscores the relevance of future research in
intermediate-energy electro-nuclear physics with polarized targets.

Specifically, in \cite{Ama99a} we computed the quasi-free $(e,e'p)$
cross section for unpolarized electrons and proton knock-out from the
$1d_{3/2}$ shell of polarized $^{39}$K leaving the residual $^{38}$Ar
in its ground state.  The analysis of the results found there as
functions of the nuclear polarization angles revealed that the nuclear
transparency indeed depends strongly on the polarization direction,
and that the FSI effects can be maximized or minimized by just
flipping the nuclear orientation. Some particular polarization
directions were found for which the FSI are practically negligible
(and thus a PWIA treatment is adequate in these cases) and others for
which the FSI even produce an enhancement of the cross section. In
conventional $(e,e'p)$ studies with unpolarized nuclei an average over
all directions is involved, this behavior is lost and the usual
reduction from the PWIA predictions, due mainly to the imaginary part
of the optical potential, is recovered.

A physical explanation of the results found in \cite{Ama99a} was also
provided, based on the PWIA and on the semi-classical concept of a
nucleon orbit. When the nucleus is polarized, the proton orbit is
oriented in a well-defined direction, and for a given value of the
missing momentum $\np$, the proton can be localized within its 
orbit (characterized by the expectation value of its position
in the semi-classical model).  In this picture of the reaction, the
electron interacts with the proton, transferring four-momentum
$Q^{\mu}=(\omega,\nq)$ and the proton leaves the nucleus after
crossing a certain amount of nuclear matter with some impact
parameter. The strength of the FSI along the
nucleon's path before it leaves the nucleus enters in such a way that in
cases where the proton is localized near the nuclear surface and
``evaporates'', immediately leaving the nucleus, the FSI are found to be small,
while in cases where it crosses the entire nucleus leaving it
by the opposite side the FSI effects are large. One can also find
cases where some components of the FSI are emphasized, for instance
when the nucleon leaves the nucleus in a direction tangent to the
surface, where it is found that the real and spin-orbit parts 
of the optical potential are the most important.  

In the case of a purely absorptive optical potential, it was possible
to fit the damping of the PWIA cross section ({\it i.e.,} explain
the nuclear transparency)
by a simple exponential factor $e^{-\rho/\lambda}$, where $\rho$ is
the computed length of the nucleon's trajectory within the nucleus and
the free parameter $\lambda$ is the nucleon mean free path.  In the
case of unpolarized nuclei other studies \cite{Bia96} have treated the
damping of the $(e,e'p)$ cross section in the lowest-order eikonal
approximation or Glauber model for high proton energies.  However,
since an average over angular orientations is implied in this case, the
position of the proton cannot be fixed.

The studies performed in \cite{Ama99a} revealed the fine control over
the FSI that can be attained by varying the nuclear
polarization. Having at hand a very intuitive physical picture of the
reaction where the reaction mechanism is expressed in geometrical
terms is clearly expected to be very helpful as guidance in
undertaking future experiments with polarized nuclei.  We wish to
emphasize the importance of performing this kind of theoretical
analysis where the behavior of the observables is explored over the
full range of polarization angles.

Very few studies exist in the literature involving exclusive reactions
with polarized medium-weight nuclei; and most of these were made only
in PWIA, as they were motivated by studies of selected aspects of the
problem, specifically, of relativistic off-shell effects in the
current \cite{Cab93} and of nuclear deformation effects in the context
of the Nilsson model \cite{Cab94,Cab95}. The two previous analyses of
the reaction that did include FSI in studies of polarization
observables with medium-weight nuclei \cite{Bof88,Gar95} presented
exploratory calculations of the cross section and/or response
functions only for a few nuclear orientations (the three usual ones
referred to the reaction plane: longitudinal, sideways and normal). In
the case of electro-nuclear reactions with light polarized targets,
the situation is different and very elaborated models already exist,
such as that of \cite{Are92} involving polarized deuterium.

In the work undertaken in \cite{Ama99a} only the simplest case of
unpolarized electrons scattered by polarized targets was
considered. In the present study we consider the more complex
situation in which the electron beam is also polarized and where a new
observable, the electron helicity-dependent (HD) cross section, comes
into play.  Guided by our previous studies here we analyze the
behavior of this observable for the full range of nuclear polarization
angles and the impact of FSI within the DWIA model for the reaction.
The results are interpreted in terms of the above-mentioned
semi-classical model, which we shall see is still valid for this
observable.  However, new physical aspects of the reaction appear in
this case: indeed not only is the expected position of the proton
within the orbit relevant, but also its spin direction comes into
play. In fact, it is well-known \cite{Ama96b,Cab93,Cab94} that, while
in PWIA the cross section for unpolarized electrons is proportional to
the spin-scalar momentum distribution of the nuclear overlap function,
the HD cross section is proportional to the {\em spin-vector} momentum
distribution. As we shall see, this can be interpreted as the nucleon
spin distribution in the orbit.

One of the goals of this paper is to provide an understanding of new
symmetry properties of the spin observables with respect to certain
class of rotations of the polarization vector that are seen in our
results.  The obvious symmetry of the observables under inversion of
the polarization vector that appears in PWIA is broken when FSI are
included. In contrast, the symmetry that we shall put in evidence is
valid even in the presence of FSI, making it a more fundamental
property of spin observables.  This symmetry occurs between nuclear
orientations that are located symmetrically with respect to the
reaction plane.  In the case of the HD cross section, it appears only
for in-plane emission and produces a change of sign, hence averaging
to zero in the case of unpolarized nuclei. For out-of-plane emission
we shall show that the symmetry of the HD cross section breaks down,
on the average yielding a non-null result for the fifth response
function for unpolarized nuclei.  The origin of this symmetry can be
explained using the semi-classical model by exploring the symmetries
of nucleon position and spin projections for different nuclear
polarizations. In particular, we shall show that the symmetry is still
present in the separate response functions even for out-of-plane
emission. The reason why it is broken in this case for the HD cross
section lies in the different behaviors found for the $T'$ and $TL'$
responses: while the $T'$ changes sign, the $TL'$ does not for
out-of-plane emission.  Since the HD cross section is a linear
combinations of these two responses, the symmetry is not preserved in
this observable.

Understanding the HD cross section has implications for the fifth
response function. In \cite{Bia97} this observable was analyzed for
proton knock-out from a $s_{1/2}$ shell assuming for the scattering
state a plane wave with complex momentum. The results were interpreted
in terms of the asymmetry between protons emitted above and below the
scattering plane and the different paths followed by the nucleon on
its way out of the nucleus. In this simple case no spin-orbit FSI
effects were included; these were discussed in \cite{Jes99}. In the
present work we consider proton knock-out from the $d_{3/2}$ shell of
$^{39}$K, which more typical, and the spin-orbit interaction is
included in the FSI. Since in our model the spin direction of the
proton is known as well as its orbital angular momentum, we are able
to show geometrically how the symmetry of the HD cross section is
preserved in the presence of a spin-orbit interaction.
 
After a short review of the DWIA model, in sect.~2 we show results for
the HD cross section over the full range of polarization directions
and for in-plane and out-of-plane proton emission, as well as for the
HD response functions, to reveal the symmetry properties of these
observables.  In sect.~3 we focus on the semi-classical model
description of the reaction and compute the expected values of
position and spin of the initial proton at a fixed value of missing
momentum for a full range of the polarization angles in order to
explain the DWIA results in a geometrical picture, giving an
analytical proof of the symmetry properties in the Appendix.  In
sect.~4 we summarize our conclusions.

\section{The DWIA model and results for the 
         helicity-dependent cross section}

The general formalism for exclusive electron scattering with polarized
beam and target can be found in \cite{Ras89}.  In this section we
summarize some of the results that are essential for our present
discussions. See \cite{Ama98a,Ama99a} for further details of our
model.

We consider an electron with four-momentum
$K^{\mu}=(\epsilon_e,\nk_e)$ and helicity $h$ which is scattered
through an angle $\theta_e$ by a nucleus, transferring four-momentum
$Q^{\mu}=(\omega,\nq)$.  We assume the extreme relativistic limit for
the electron $\epsilon_e\gg m_e$ and the plane-wave Born approximation
for the electron scattering amplitude.  We work in a coordinate system
where the $x$-$z$ plane is the scattering plane, with the $z$-axis in
the direction $\nq$ and the $x$ axis defined by the transverse
component of the electron momentum $\nk_e$.

After the scattering, a proton with energy $E'$ is detected in the
direction $\hp'=(\theta',\phi')$ in coincidence with the electron.  In
this work the momentum of the ejectile is determined assuming
relativistic kinematics $E'{}^2-\np'{}^2=M^2$, where $M$ is the
nucleon mass.  We work in the lab. frame where the target nucleus is at rest
in its ground state. We assume that it has total spin $J_i$
and is 100\% polarized in the direction $\nOmega^*$ defined by the
polar and azimuthal angles $\Omega^*=(\theta^*,\phi^*)$, measured in
the above coordinate system: 
\begin{equation}
|A(\Omega^*)\rangle = |J_iJ_i(\Omega^*)\rangle .
\end{equation}
The unobserved residual nucleus is assumed to be left in a discrete state 
$|B\rangle$ and we neglect its
recoil. The corresponding coincidence cross section can be written as
\begin{equation}
\frac{d\sigma}{d\epsilon'_e d\Omega'_e d\hp'}
= \Sigma+h\Delta ,
\end{equation}
where $\Sigma$ is the cross section for unpolarized electrons
\begin{equation}
\Sigma=
\sigma_{Mott}
(v_L{\cal R}^L+v_T{\cal R}^T+v_{TL}{\cal R}^{TL}+v_{TT}{\cal R}^{TT})
\end{equation}
and $\Delta$, the observable  we are interested in here,  
is the helicity-dependent cross section, given by
\begin{equation}\label{Delta}
\Delta = \sigma_{Mott}(v_{T'}{\cal R}^{T'}+v_{TL'}{\cal R}^{TL'}).
\end{equation}
The quantities $v_K$ and $v_{K'}$ are electron kinematical factors
coming from the leptonic tensor \cite{Ras89}. The exclusive response
functions ${\cal R}^K$, ${\cal R}^{K'}$ are specific components of the
hadronic tensor
\begin{equation}
W^{\mu\nu} = \sum_{m_s,M_B}
\langle \np' m_s,B|J^{\mu}(\nq,\omega)|A\rangle^*
\langle \np' m_s,B|J^{\nu}(\nq,\omega)|A\rangle
\end{equation}
containing the transition matrix elements of the nuclear electromagnetic
current operator $J^{\mu}(\nq,\omega)$.  The final
(unpolarized) state is defined by the asymptotic boundary condition
of a nucleon with momentum $\np'$ and spin projection $m_s$,
together with a daughter nucleus $|B\rangle = |J_B M_B\rangle$.
We sum over final, undetected magnetic quantum numbers.

The relevant response functions in this work are the helicity-dependent 
ones, defined by
\begin{eqnarray}
{\cal R}^{T'} &=& i(W^{xy}-W^{yx}) \equiv W^{T'} \label{T'}\\
{\cal R}^{TL'} &=& i\sqrt{2}(W^{0y}-W^{y0}) 
\equiv W^{TL'}\sin\phi'+\tilde{W}^{TL'}\cos\phi', \label{TL'}
\end{eqnarray}
where we have extracted the explicit dependence on the azimuthal
emission angle $\phi'$.  The response functions $W^{T'}$, $W^{TL'}$
and $\tilde{W}^{TL'}$ depend on $q$, $\omega$, $\theta'$, $\theta^*$
and $\Delta\phi\equiv\phi'-\phi^*$.  In this work we fix the values of
the energy-momentum transfer around the quasielastic peak, $p'\sim
q$, where the impulse approximation used here is expected to to work
well. We present results for the HD cross section $\Delta$ as a
function of the missing momentum $p$, defined as $\np=\np'-\nq$, for
a range of values of the angles $\phi$, $\theta^*$ and $\Delta\phi$.

For the electromagnetic operators we consider a new expansion 
of the relativistic on-shell current 
in powers of $\np/M$, with $\np$ the initial nucleon 
momentum, eliminating the final momentum $\np'$ using momentum
conservation \cite{Ama96a,Ama98b}. 
This is a very convenient expansion for
intermediate-to-high energy quasi-free processes where 
the initial nucleon is in a bound
state (and therefore its momentum is small enough so that a
first-order expansion works well), but the final momentum is large
(comparable with $q$) and hence the  traditional  expansions in 
$\np'/M$ are bound to fail. 
Under this expansion, the charge and transverse current operators in
momentum space are
given by 
\begin{eqnarray}
\rho(\np',\np) &=&
\rho_c+i\rho_{so}(\cos\phi\sigma_y-\sin\phi\sigma_x)\delta
\label{sn-current-0}\\
J^x(\np',\np) &=&
iJ_m\sigma_y + J_c\delta\cos\phi
\label{sn-current-x}
\\
J^y(\np',\np) &=&
-iJ_m\sigma_x + J_c\delta\sin\phi ,
\label{sn-current-y}
\end{eqnarray}
where $\delta\equiv \eta\sin\theta$, $\eta=p/M$, and 
the angles $(\theta,\phi)$
define the direction of the initial momentum $\np$.
Note the spin-dependence explicit in  the Pauli matrices.
Finally, using our previous work the coefficients $\rho_c$ (charge), $\rho_{so}$
(spin-orbit), $J_m$ (magnetization), and $J_c$ (convection)  
are given by
\begin{eqnarray}
\rho_c = \frac{\kappa}{\sqrt{\tau}}G_E,
&&
\rho_{so} = \frac{2G_M-G_E}{\sqrt{1+\tau}}\frac{\kappa}{2},
\\
J_m = \sqrt{\tau}G_M, 
&&
J_c = \frac{\sqrt{\tau}}{\kappa}G_E.
\end{eqnarray}
Here, as in past work, we use the dimensionless variables $\kappa=q/2M$ and
$\tau=(q^2-\omega^2)/4M^2$, and $G_E$ and $G_M$ are the electric and
magnetic form factors of the nucleon, respectively, for which we use
the Galster parameterization \cite{Gal71}.

The inclusive electromagnetic responses obtained in the present
model for the current were tested against the relativistic Fermi gas model
in \cite{Ama96a}, while in \cite{Udi99} our present DWIA
model of unpolarized $(e,e'p)$ reaction was compared with a 
fully-relativistic model for $|Q^2|=0.8$ (GeV/c)$^2$. In both cases for
quasi-free kinematics the expansion procedure was shown to be robust.

\begin{figure}[hptb]
\begin{center}
\leavevmode
\def\epsfsize#1#2{0.8#1}
\epsfbox[100 160 500 781]{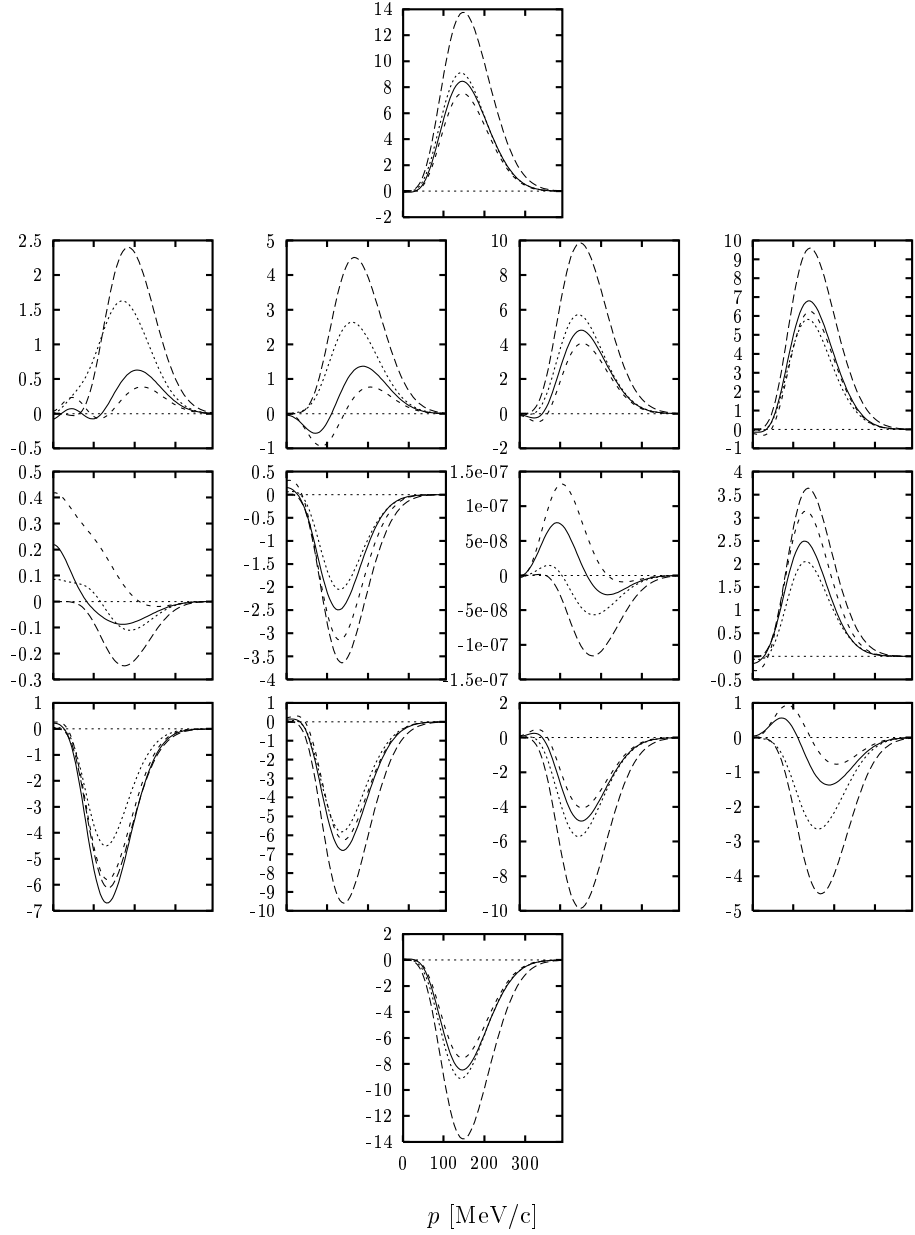}
\end{center}
\caption{Helicity-dependent cross section for in-plane emission, as a
function of the missing momentum. The various panels correspond to
different pairs of polarization angles $(\theta^*,\Delta\phi)$. From up
to down $\theta^*=0,45,90,135$ and 180$^{\rm o}$; from left to right
$\Delta\phi=0,45,90$ and 135$^{\rm o}$. The meaning of the curves is
the following:  solid: DWIA; long-dashed: PWIA;
dotted: DWIA but with just the imaginary part of the central optical
potential; short-dashed: DWIA without spin-orbit contributions }
\end{figure}

\begin{figure}[hptb]
\begin{center}
\leavevmode
\def\epsfsize#1#2{0.8#1}
\epsfbox[100 380 500 781]{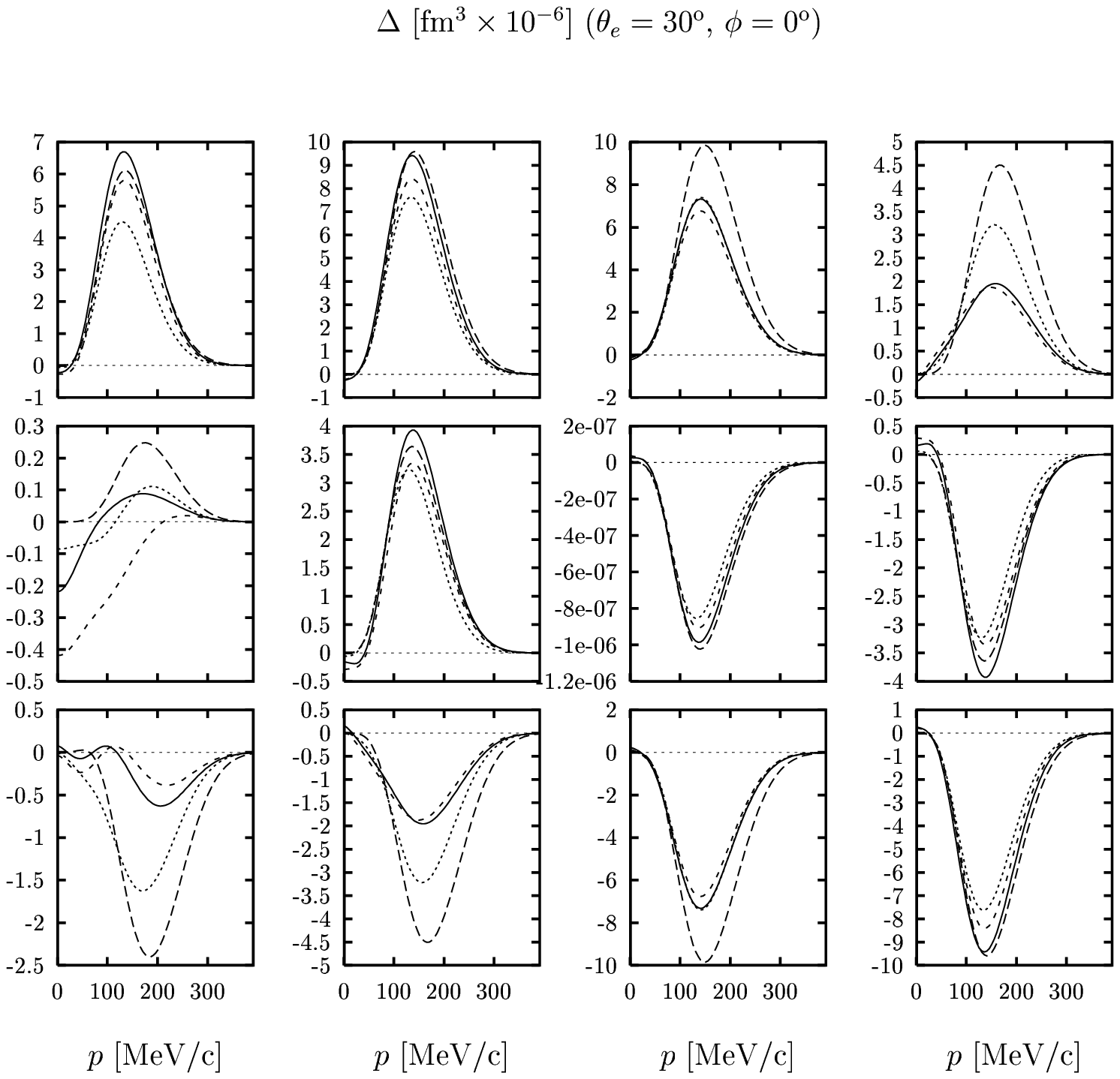}
\end{center}
\caption{
Helicity-dependent cross section for in-plane emission, as a
function of the missing momentum. The various panels correspond to
different pairs of polarization angles $(\theta^*,\Delta\phi)$. From up
to down $\theta^*=45,90$ and 135$^{\rm o}$; from left to right
$\Delta\phi=180,225,270$ and 315$^{\rm o}$. The meaning of the curves is
the same as in Fig.~1.
}
\end{figure}

In Figs.~1--4 we present results for the polarized HD cross section
$\Delta$ for the reaction $^{39}$K$(e,e'p)\,^{38}$Ar$_{g.s.}$ using
a range of values of the angles $\phi$, $\theta^*$ and $\Delta\phi$.
The reason for choosing the $^{39}$K nucleus as target
is twofold: first, being the nucleus of an alkali atom potassium is a
good candidate to be polarized in experimental studies. Second, it is a
medium-weight, closed-shell-minus-one nucleus and therefore the shell model is
expected to be adequate for describing its nuclear structure in such a first
exploratory study. Hence in our calculation 
the $^{39}$K target is described as a proton hole in the
$1d_{3/2}$ shell of the $^{40}$Ca core, while the residual $^{38}$Ar
nucleus is described as two holes in the $1d_{3/2}$ shell, coupled to
zero total angular momentum.  Single-particle wave functions in a
Woods-Saxon mean potential are used for the hole states, with
parameters taken from \cite{Ama94}.  The ejected proton wave 
function is obtained by solving the
Schr\"odinger equation in an optical potential for which we employ the
Schwandt parameterization \cite{Sch82}. More details on these aspects
of our model are given in \cite{Ama98a,Ama96b}.

In our calculations we have fixed the momentum transfer to 
$q=500$ MeV/c and the energy transfer to $\omega=133.5$ MeV, corresponding to the
quasielastic peak, and the electron scattering angle to 
$\theta_e=30^{\rm o}$. For this choice of kinematics, the 
relevant leptonic factors take on the values
$v_{T'}= 0.268$ and $v_{TL'}= -0.176$. The kinetic energy of the 
ejected proton is 124.8 MeV, while its momentum is $p'=499.7$ MeV/c.

In Fig.~1 the HD cross section $\Delta$ is shown for in-plane
emission, $\phi=0^{\rm o}$ as a function of the missing momentum.  
The various panels correspond to different values of the 
polarization angles in multiples of 45$^{\rm o}$; 
from up to down $\theta^*=0,45,90,135$ and 180$^{\rm o}$;
from left to right $\Delta\phi=0,45,90$ and 135$^{\rm o}$. 
Note that in this case $\Delta\phi=-\phi^*$. Hence the polarization
vectors in the 14 panels of
Fig.~1 span a hemisphere. The other half of the sphere is
presented in Fig.~2, where now $\theta^*=45,90,135^{\rm o}$ and
$\Delta\phi=180,225,270,315^{\rm o}$.
In each panel of Figs.~1--2 we show four curves corresponding to
different treatments of the FSI. Solid lines correspond to the 
full optical potential, while long-dashed lines have been obtained
without FSI, {\it i.e.,} setting the optical potential equal to zero, and
thus they coincide with the factorized PWIA. The dotted lines include
only the imaginary part of the central optical potential and
correspond to a purely absorptive model of the interaction. Finally, the
short-dashed lines include as well the real part of the central optical
potential. Neither the dotted nor the short-dashed lines include the
spin-orbit part of the potential, which is only considered in the
case of the solid lines. 
Looking at the various curves in the figures we get a
feeling for the effects of the separate ingredients in the FSI 
for the full range of values of the polarization angles and find a wide
variety of situations, going from large FSI, for instance 
for $\theta^*=45^{\rm o}$ and $\Delta\phi=0^{\rm o}$ (Fig.~1), to small FSI
for  $\theta^*=45^{\rm o}$ and $\Delta\phi=180^{\rm o}$ (Fig.~2). 
Such observations were explained in \cite{Ama99a} for the case of the 
helicity-independent cross section $\Sigma$. They arise from the fact
that the length of the nucleon's path inside the nucleus 
is larger in the former case than in the latter. In the next section
we will come back to the generalization to HD observables and again
see how the semi-classical model may be used to shed light on the results.

Another point to emphasize here is the utility of having a plot of $\Delta$
for the full range of polarization angles as in Figs.~1--2 where we can study the
symmetries of this observable by direct inspection.
First we note that in going from one polarization to the opposite, the
PWIA HD cross section changes its sign (long-dashed lines), but that this is
not in general the case when the FSI are turned on. Compare, for
instance the polarization $\theta^*=135^{\rm o}$, $\Delta\phi=45^{\rm
o}$ (Fig.~1), with the opposite $\theta^*=45^{\rm o}$,
$\Delta\phi=225^{\rm o}$ (Fig.~2).  In the first case the FSI produce a 30\%
decrease of the strength (solid lines), while in the second the total
FSI effects are negligible.  We also note from Figs.~1--2 that only for
opposing pairs having $\Delta\phi=0^{\rm o}$ or  $180^{\rm o}$ does the
HD cross section cancellation occur.
Therefore the fact that the HD cross section for
unpolarized nuclei is zero for $\phi=0^{\rm o}$ even in presence of FSI is not
due to cancellations between {\em opposite} polarizations when taking the
average over all angles.

\begin{figure}[hptb]
\begin{center}
\leavevmode
\def\epsfsize#1#2{0.8#1}
\epsfbox[100 160 500 781]{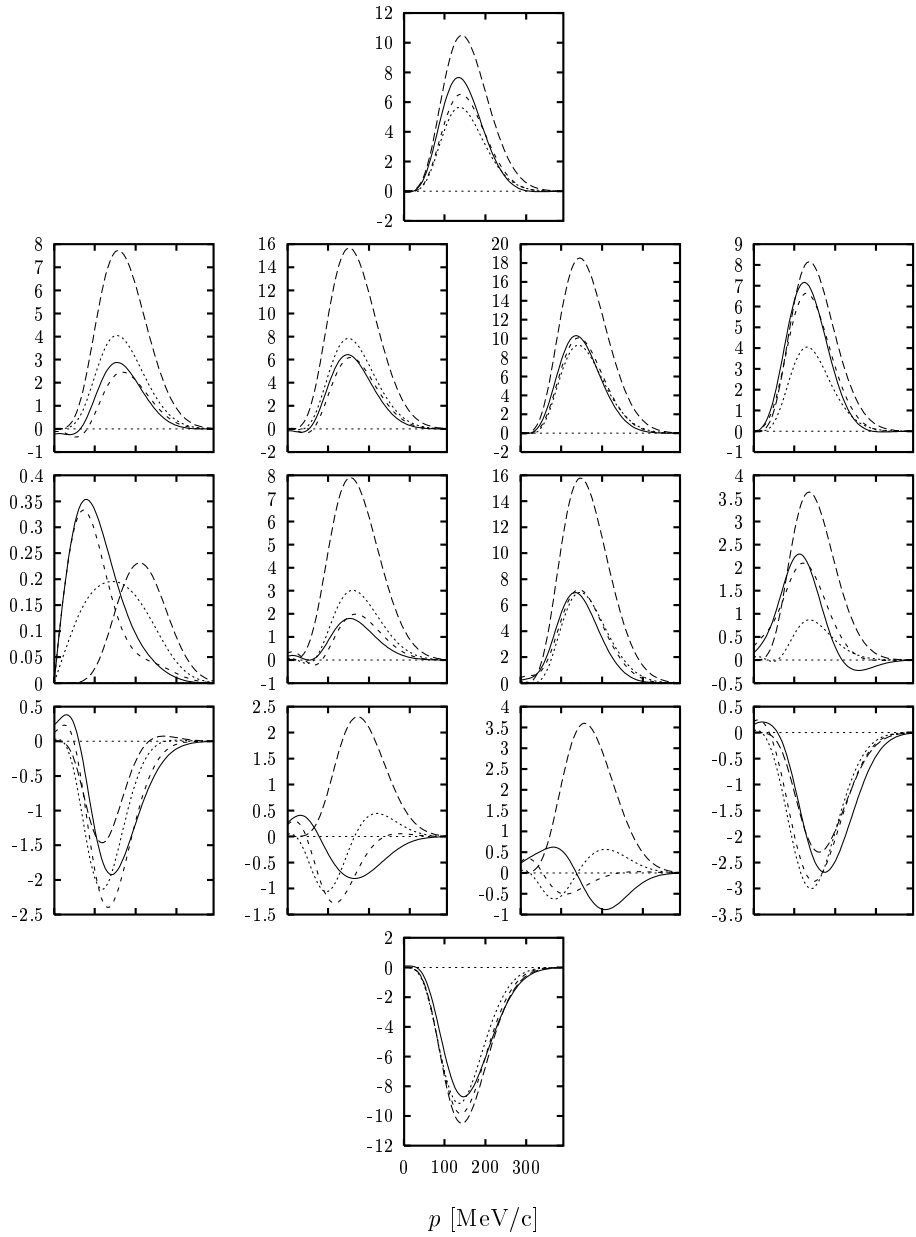}
\end{center}
\caption{
The same as Fig.~1, but now for $\phi=90^{\rm o}$.
}
\end{figure}

\begin{figure}[hptb]
\begin{center}
\leavevmode
\def\epsfsize#1#2{0.8#1}
\epsfbox[100 380 500 781]{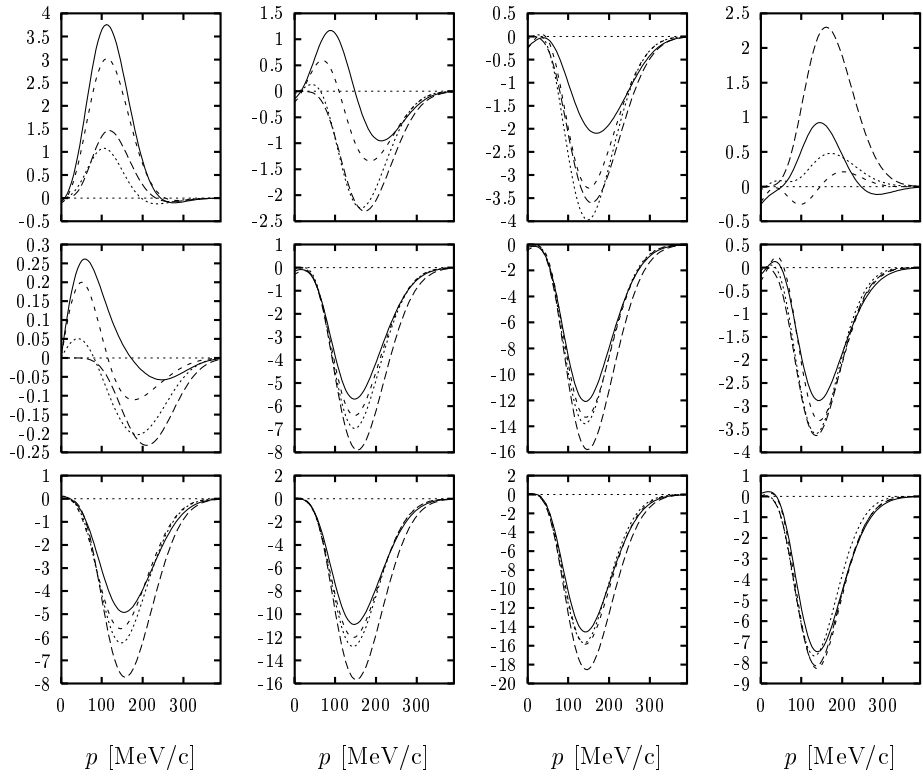}
\end{center}
\caption{
The same as Fig.~2, but now for $\phi=90^{\rm o}$.
}
\end{figure}

Thus the following question arises: how do the cancellations occur in
this observable if not for the behavior for opposite polarizations?
The answer is again contained in the results of Figs.~1--2. Upon
closer examination a new symmetry in the HD cross section becomes
apparent.  But this time it is exact even in presence of the FSI.  In
fact, if we take the origin of coordinates in the
$(\theta,\Delta\phi)$ plane as the point $(90^{\rm o},90^{\rm o})$,
represented in Fig.~1 as the panel in the third column and third row,
then we see that $\Delta$ takes opposite values for polarization
angles that are opposite with respect to this central point.  Compare
for instance the pairs of polarizations $(45^{\rm o},45^{\rm o})$ and
$(135^{\rm o},135^{\rm o})$ or $(45^{\rm o},90^{\rm o})$ and
$(135^{\rm o},90^{\rm o})$.  The same kind of ``nodal'' symmetry is
observed in Fig.~2 when we take as central point (or node) the panel
$(\theta,\Delta \phi)=(90^{\rm o},270^{\rm o})$.  In this way it
appears that, when averaging over polarization angles, there will be
cancellations between pairs of polarizations located symmetrically
with respect to any of the two nodal panels and a zero value of
$\Delta$ for unpolarized nuclei is expected.  Note that the two nodal
points correspond to polarization vectors that lie perpendicular to
the reaction plane (or to the scattering plane in this case). 
Were the model calculations to be exact, we would expect to find
that the HD cross section corresponding to the two nodes is
exactly zero, as would be implied by the nodal symmetry. In fact,
we see that it is very small in the two cases (of the order of 
$10^{-7}$--$10^{-6}$ in the units of the figures) --- these values 
are within the numerical error of the multipole expansion in our 
calculation and hence they are compatible with zero. We have checked 
that the cited symmetry between various pairs of polarizations is 
always satisfied within this uncertainty and hence we conjecture
that this nodal symmetry is an exact one. 
The same symmetry is also present in the
helicity-independent cross section studied in \cite{Ama99a}, although
this fact was not mentioned in that reference.  The goal of the next
section will be to explain its origin within the semi-classical model
for the nucleon orbit using geometrical arguments.

Up to this point we have discussed only results for in-plane emission
kinematics ($\phi=0^{\rm o}$).  Now in Figs.~3--4 we show results for
$\phi=90^{\rm o}$, where the reaction plane is perpendicular to the
scattering one. The meaning of lines and of the various panels is
the same as in Figs.~1--2, corresponding to angles $\theta^*$ and
$\Delta\phi$ spanning all directions, but note that
$\Delta\phi=90^{\rm o}-\phi^*$ is now minus the azimuthal polarization
angle measured with respect to the reaction plane.  Looking at the
figures we note again that the PWIA results (long-dashed lines) are
opposite for opposite polarizations and hence a result of
zero for this observable is obtained, corresponding to the expected
result for unpolarized nuclei when there are no FSI.  In the presence of
FSI the different paths followed by the proton for opposite
polarizations are reflected again in different effects due to
FSI. These are larger in general for the polarizations of Fig.~3 than
for those of Fig.~4. A wide range of effects can be found, from
situations where FSI produce a very large decrease of $\Delta$, for
$(\theta^*,\Delta\phi)=(90^{\rm o},45^{\rm o})$ (Fig.~3), 
even a change of sign for $(\theta^*,\Delta\phi)=(135^{\rm o},45^{\rm
  o})$ (Fig.~3), passing cases with small
effects for $(\theta^*,\Delta\phi)=(135^{\rm o},315^{\rm o})$
(Fig.~4), to cases with a large {\em increase} due to FSI for 
$(\theta^*,\Delta\phi)=(45^{\rm o},180^{\rm o})$ (Fig.~4).  

A significant difference with respect to the $\phi=0^{\rm o}$
case is that the nodal symmetry with respect to the $(90^{\rm o},90^{\rm o})$
polarization is not found here. No other symmetry is observed and, as
a consequence, there will be a non-zero result predicted for this
observable if the nucleus is unpolarized, as expected, and this in
the end will give rise to the fifth response function which can be
measured by performing out-of-plane measurements with polarized
electrons, but unpolarized nuclei.
 
\begin{figure}[hptb]
\begin{center}
\leavevmode
\def\epsfsize#1#2{0.8#1}
\epsfbox[100 160 500 781]{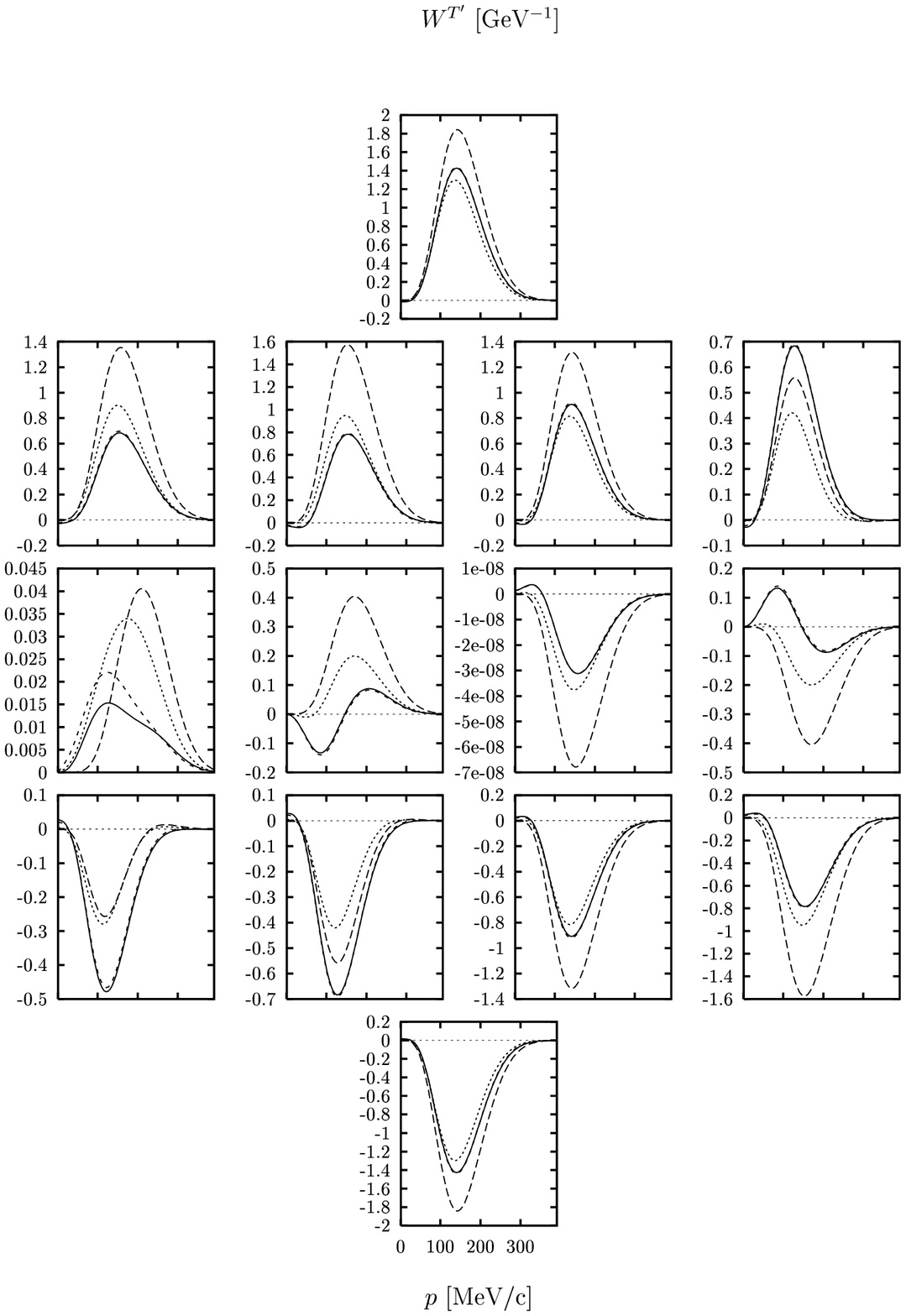}
\end{center}
\caption{
Response function $W^{T'}$ computed for a range of nuclear
polarization angles $(\theta^*,\Delta\phi)$.
  The meaning of the panels and curves is the same as in
Fig.~1.
}
\end{figure}

\begin{figure}[hptb]
\begin{center}
\leavevmode
\def\epsfsize#1#2{0.8#1}
\epsfbox[100 380 500 781]{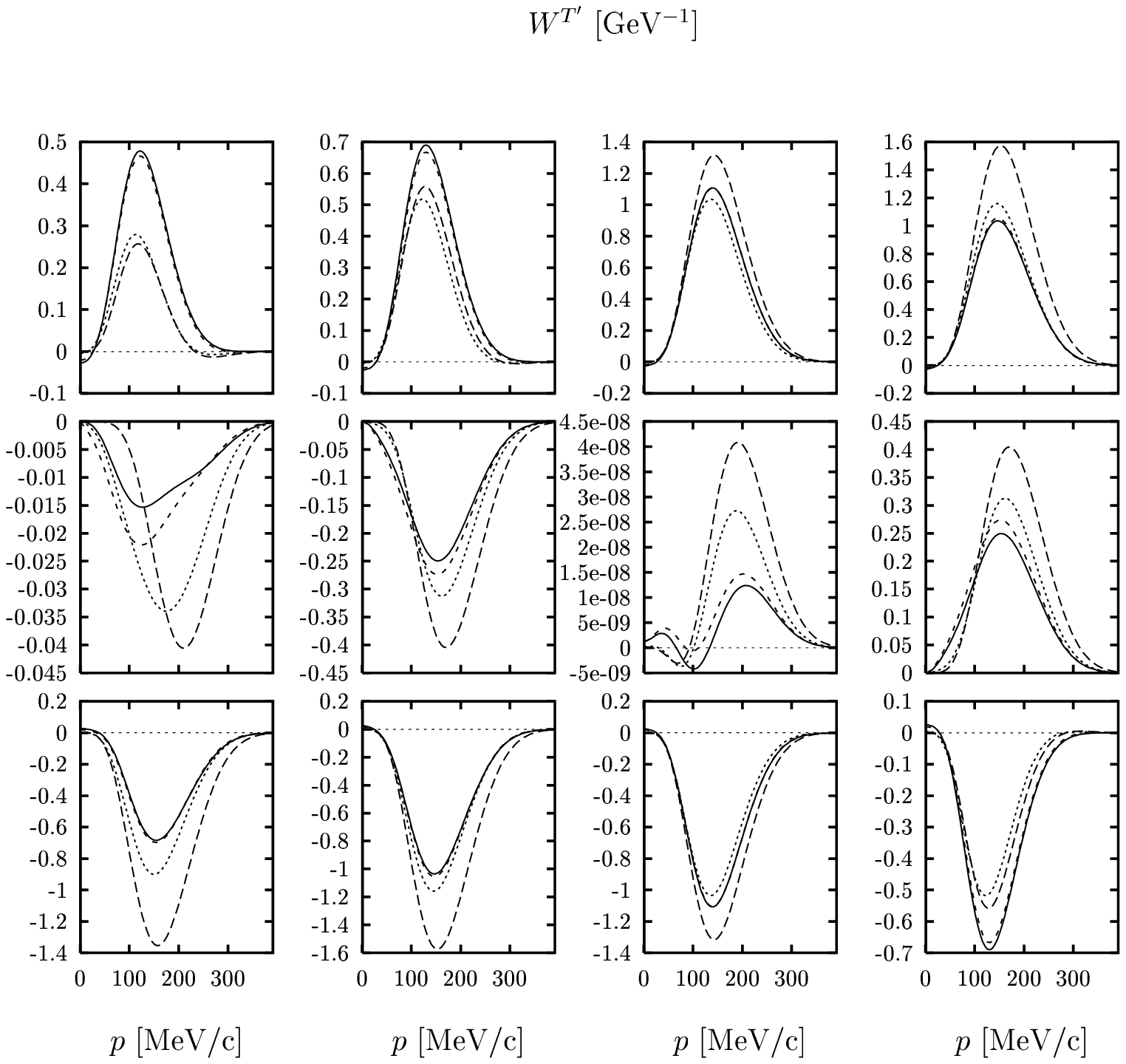}
\end{center}
\caption{ Response function $W^{T'}$ computed for a range of nuclear
polarization angles $(\theta^*,\Delta\phi)$.  The meaning of the
panels and curves is the same as in Fig.~2.  }
\end{figure}

\begin{figure}[hptb]
\begin{center}
\leavevmode
\def\epsfsize#1#2{0.8#1}
\epsfbox[100 160 500 781]{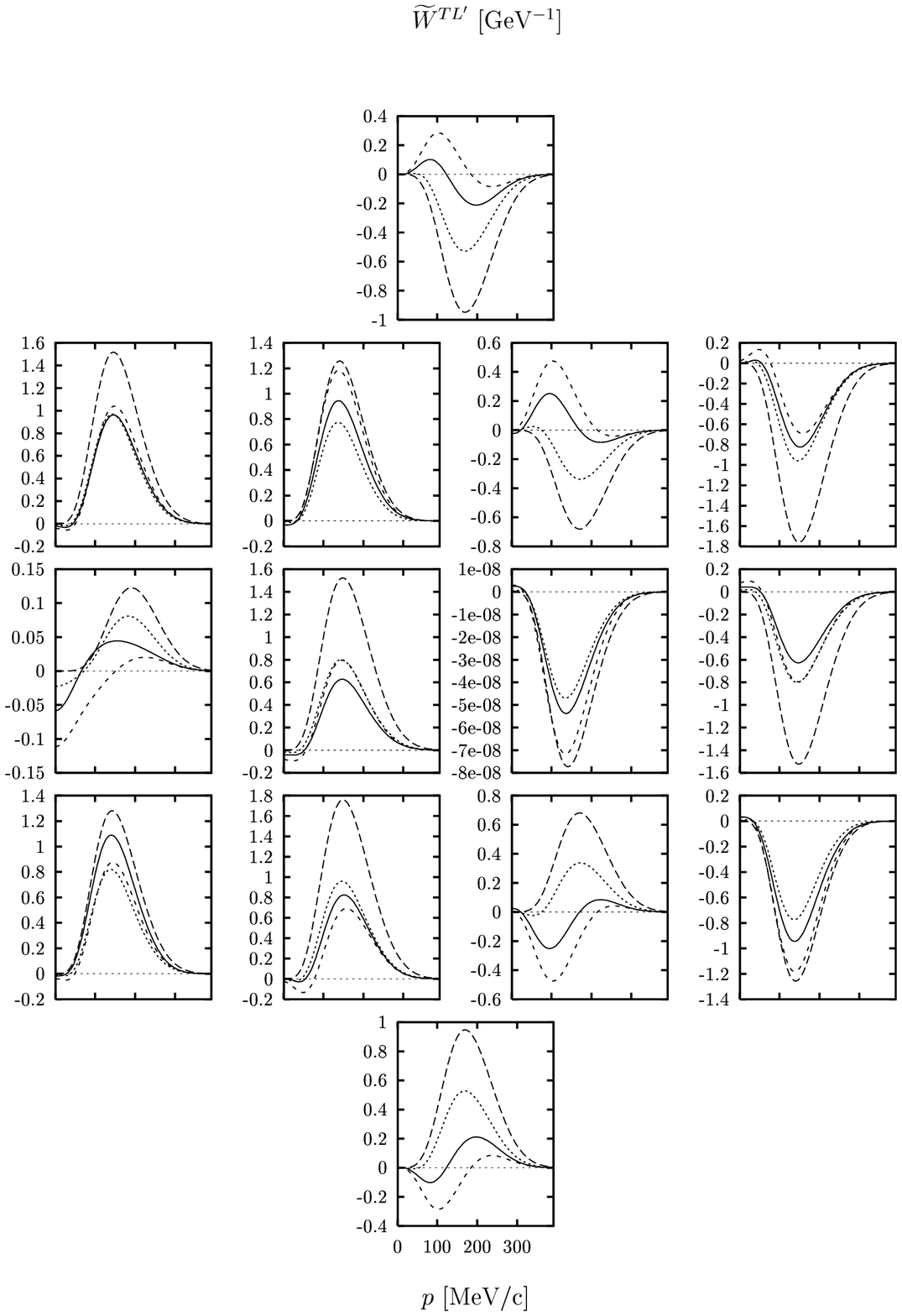}
\end{center}
\caption{ Response function $\widetilde{W}^{TL'}$ computed for
a range of nuclear polarization angles $(\theta^*,\Delta\phi)$.  The
meaning of the panels and curves is the same as in Fig.~1.  }
\end{figure}

\begin{figure}[hptb]
\begin{center}
\leavevmode
\def\epsfsize#1#2{0.8#1}
\epsfbox[100 380 500 781]{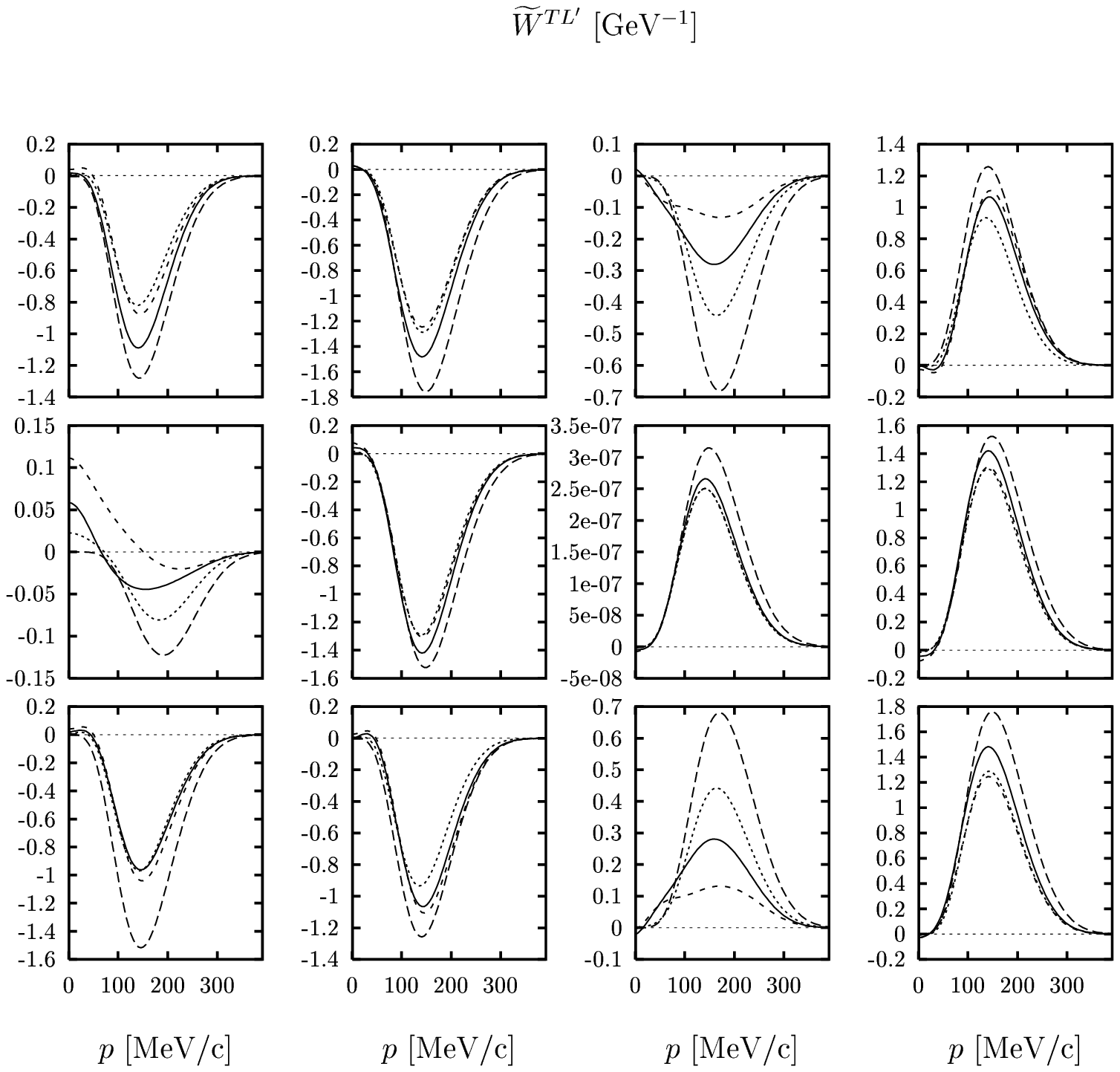}
\end{center}
\caption{ Response function $\widetilde{W}^{TL'}$ computed for
a range of nuclear polarization angles $(\theta^*,\Delta\phi)$.  The
meaning of the panels and curves is the same as in Fig.~2.  }
\end{figure}

\begin{figure}[hptb]
\begin{center}
\leavevmode
\def\epsfsize#1#2{0.8#1}
\epsfbox[100 160 500 781]{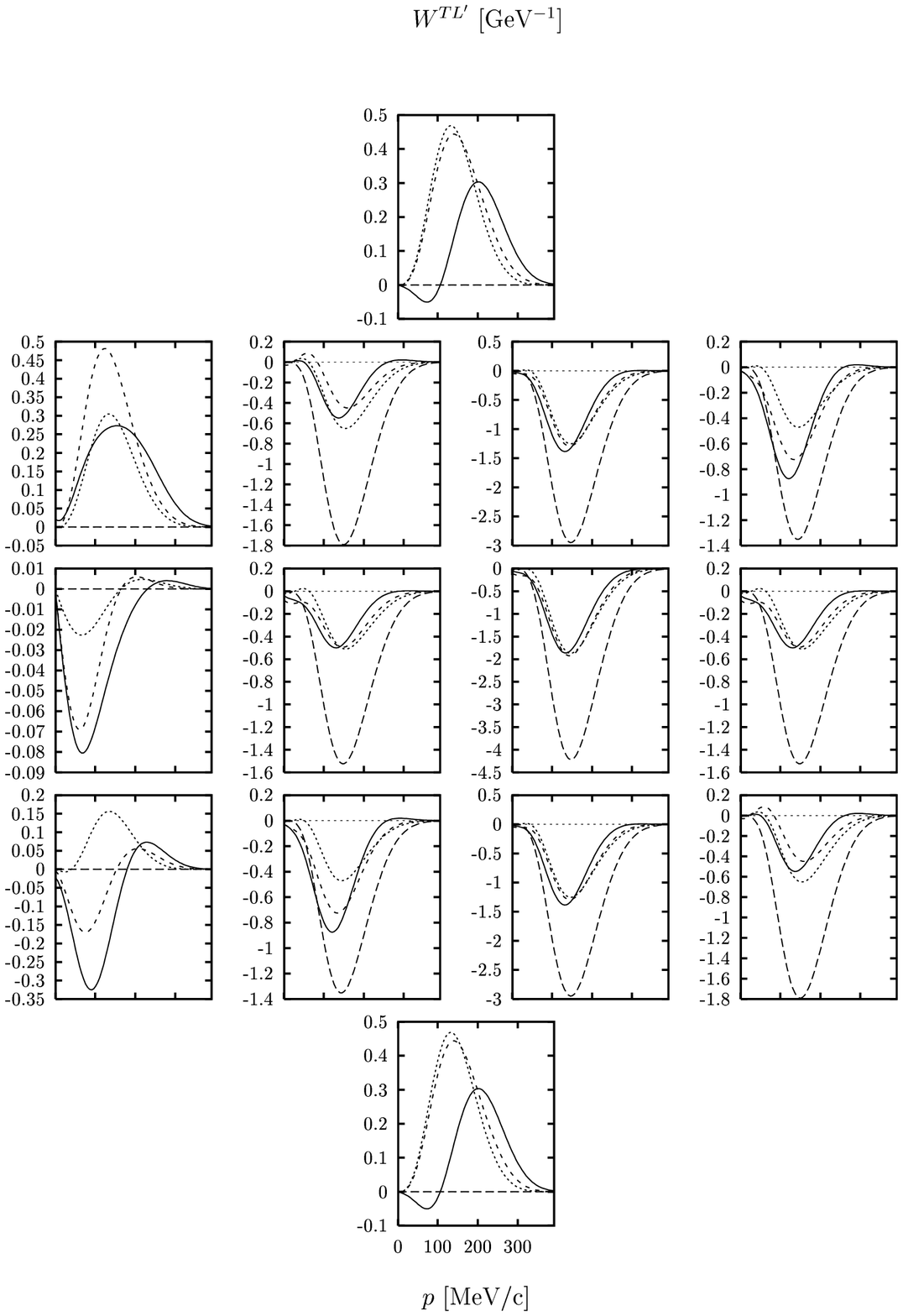}
\end{center}
\caption{ Response function $W^{TL'}$ computed for a range of nuclear
polarization angles $(\theta^*,\Delta\phi)$.  The meaning of the
panels and curves is the same as in Fig.~1.  }
\end{figure}

\begin{figure}[hptb]
\begin{center}
\leavevmode
\def\epsfsize#1#2{0.8#1}
\epsfbox[100 380 500 781]{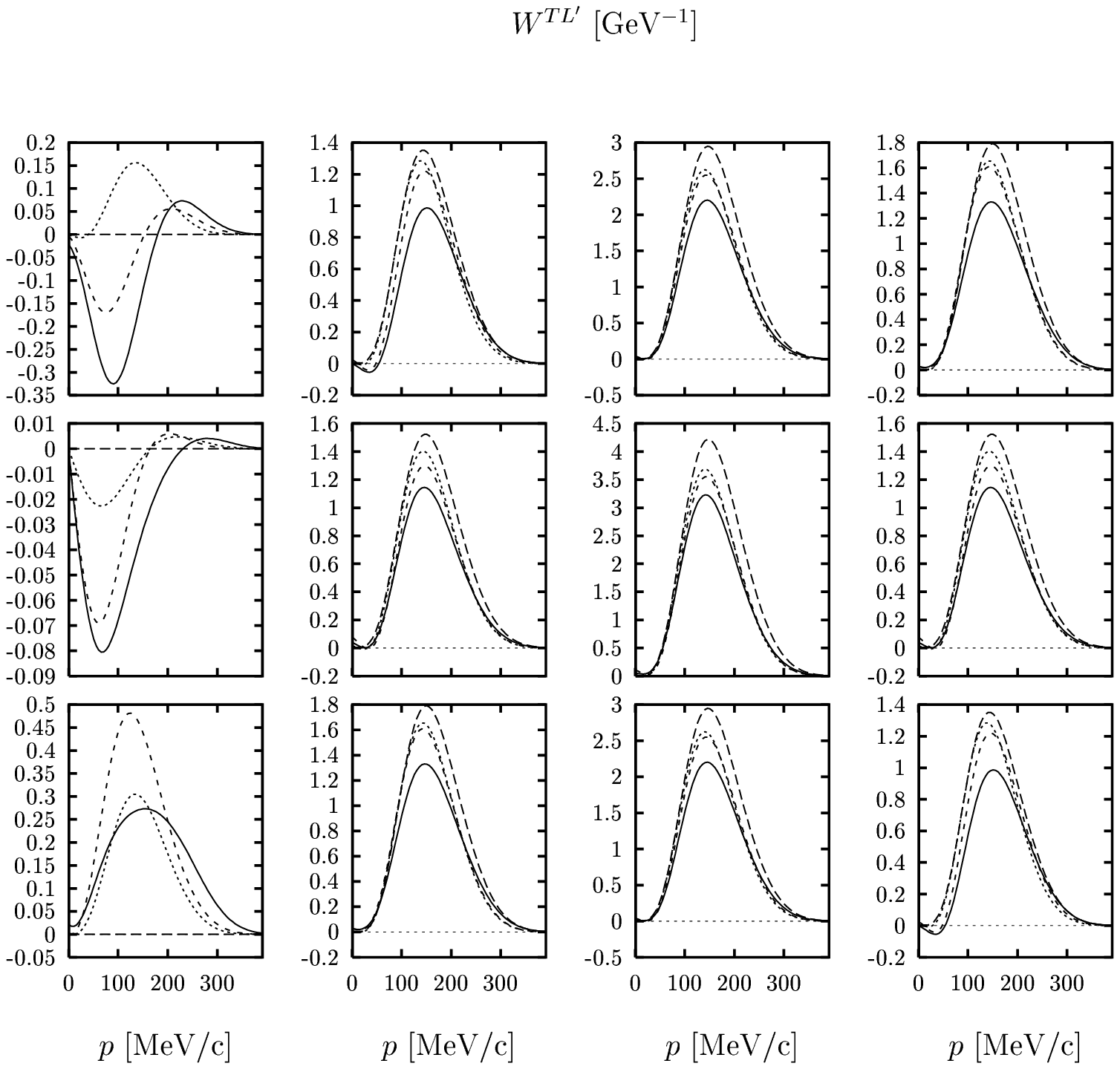}
\end{center}
\caption{ Response function $\widetilde{W}^{TL'}$ computed for
a range of nuclear polarization angles $(\theta^*,\Delta\phi)$.  The
meaning of the panels and curves is the same as in Fig.~2.  }
\end{figure}

Deeper insight into the reasons why the nodal symmetry is broken for
out-of-plane emission can be obtained if we study the behavior of the
three response functions $W^{T'}$, $W^{TL'}$ and $\widetilde{W}^{TL'}$
which contribute to $\Delta$ (see
Eqs.~(\ref{Delta},\ref{T'},\ref{TL'})).  These are shown in
Figs.~5--10, again for the full set of values of polarization angles
$(\theta^*,\Delta\phi)$.  Note that these responses do not depend on
$\phi$. The $T'$ response gives the same contribution to $\Delta$ for
any value of $\phi$; on the other hand, in the case of the $TL'$
response for $\phi=0^{\rm o}$ ($90^{\rm o}$) only the
$\widetilde{W}^{TL'}$ ($W^{TL'}$) response contributes.  Since for
$\phi=0^{\rm o}$ we found nodal symmetry in the HD cross section, this
symmetry should also be present in the two response functions which
contribute to this observable, and in fact, we see in Figs.~5--6 and
7--8 that both $W^{T'}$ and $\widetilde{W}^{TL'}$ responses show this
symmetry, changing sign for opposite polarizations with respect the
nodal polarization angles $(\theta^*,\Delta\phi)=(90^{\rm o},90^{\rm
o})$.  As a consequence, any linear combination of these two
responses, such as the observable $\Delta$ for $\phi=0^{\rm o}$, will
present the same nodal symmetry.  Both responses are zero (within the
numerical precision --- see above) for the nodal polarizations,
although the shape and sign of these two responses are in general
different for other polarizations, and the effects of the FSI,
especially those related to the real and spin-orbit parts of the
optical potential, are also different.

Let us now examine the behavior of the $W^{TL'}$ response shown in
Figs.~9--10. As before we note that in PWIA this response changes sign 
for opposite polarizations, but that again this is no longer true when FSI
are switched on, due to the different path traveled by nucleons in
their way across the nucleus, this being larger for the polarizations of
Fig.~9 than for those of Fig.~10. The most important difference for
this response function as compared with the others is its behavior
under polarization inversion with respect to the nodal polarization
$(\theta^*,\Delta\phi)=(90^{\rm o},90^{\rm o})$. 
This response still displays a symmetry even in presence of FSI,
yet in this case there is no change of
sign, {\it i.e.,} this response is {\em even} under the nodal inversion 
of polarization, while the $W^{T'}$ and $\widetilde{W}^{TL'}$ 
responses are {\em odd} under it. 
This behavior explains why the fifth response function for unpolarized
nuclei, obtained as an average of the results in Figs.~9--10, is not
zero. However, the fact that the nodal symmetry persists in the three
response functions, either of odd or even type, is an indication of
the fundamental nature  of this symmetry. In the case of the 
observable $\Delta$ for $\phi\ne 0^{\rm o}$, where responses with both kinds of
symmetries appear, the nodal symmetry is broken, 
since the linear combination of
odd and even functions does not display any symmetry.

In the last instance, the present results provide an alternative  
way to separate the $W^{T'}$ from the ${W}^{TL'}$ response
in the HD cross section by performing a measurement with
$\phi=90^{\rm o}$ for two different  polarizations, $\Omega_1$ and
$\Omega_2$, that are  opposite 
with respect to the nodal polarization,
and exploiting the odd and even 
symmetries of these responses. 
In fact, we just add and subtract the two values of $\Delta$ so
obtained. From Eqs.~(\ref{Delta},\ref{T'},\ref{TL'}) 
we have
\begin{eqnarray}
\Delta(\Omega_1)+\Delta(\Omega_2)
& = & 2\sigma_{Mott}v_{TL'}{W}^{TL'}(\Omega_1)
\\
\Delta(\Omega_1)-\Delta(\Omega_2)
& = & 2\sigma_{Mott}v_{T'}W^{T'}(\Omega_1).
\end{eqnarray}

\section{The semi-classical model and 
         applications to the  HD cross section}

In this section we analyze the results of the previous section 
in terms of geometrical properties of the nucleon orbit for various
nuclear polarizations and kinematics. In particular we shall see that 
the behavior observed for the response functions and HD cross section can
be characterized by two parameters, namely, the expectation value of the
nucleon position within the orbit, and the expectation value of its
spin. One of our goals is to express in geometrical terms why 
the nodal symmetry (in both versions, even and odd) is happening. 

The semi-classical model was introduced in
\cite{Ama99a} to explain the results of the $(e,e'p)$ cross
section $\Sigma$ for  polarized nuclei.
It was based on the PWIA, the 3-dimensional scalar momentum
distribution and the expected position of the nucleon in its orbit.  
For the present applications of this 
picture to the observable $\Delta$ a new element has to be added
to that model, namely the expected value of the nucleon's spin.

\subsection{PWIA responses}

A characteristic of PWIA is that the exclusive responses 
(and hence the cross section) 
factorize as the product of a
single-nucleon scalar or vector response times a scalar or vector
momentum distribution 
\begin{eqnarray}
{\cal R}^K    &=& Mp' w^K_S M_S,     \kern 1cm K=L,T,TL,TT \label{RK}\\
{\cal R}^{K'} &=& Mp' \nw^{K'}_V \cdot \nM_V, \kern 1cm K'=T',TL' \label{RK'}
\end{eqnarray}
where $M_S(\np)$ and 
$M_V(\np)$ are respectively the scalar and
vector momentum distributions. They are related to the 
partial momentum distribution spin matrix, $n(\np)$, 
for residual nucleus $|B\rangle$ through
\begin{equation}
n(\np)_{\mu\nu}\equiv \sum_{M_B} \langle B|a_{\np \nu}|A\rangle^*
\langle B|a_{\np \mu}|A\rangle  
= \frac12
(M_S(\np)\delta_{\mu\nu}+\nM_V(\np)\cdot\nsigma_{\mu\nu}) .
\end{equation}
The scalar $w_S^K$ and vector $\nw_V^K$ single-nucleon responses are
obtained from the corresponding single-nucleon current matrix
\cite{Ama96b}.  For the situation of interest here with polarized electrons,
only the vector responses contribute to the HD cross section.  In the
case of the single-nucleon current defined in 
Eqs.~(\ref{sn-current-0}--\ref{sn-current-y}) the corresponding
expressions are the following:
\begin{eqnarray}
\nw_V^{T'} 
&=&
2J_c J_m \delta(\cos\phi\,\nne_1+\sin\phi\,\nne_2)-2J_m^2\nne_3
\label{wt'}\\
\nw_V^{TL'} 
&=& 2\sqrt{2}\rho_c J_m \nne_1
+2\sqrt{2} J_c\rho_{so}\delta^2\sin\phi(\cos\phi\,\nne_2-\sin\phi\,\nne_1)
\nonumber\\
&& -2\sqrt{2}\rho_{so}J_m\delta \cos\phi\, \nne_3 ,
\label{wtl'}
\end{eqnarray}
where the unit vectors $\nne_i$ correspond to the coordinate system 
$O$ introduced above, with $\nne_3$ in the $\nq$-direction and 
$\nne_1$ in the scattering plane (in the direction of $\nk_{eT}$), 
while $\nne_2$ is perpendicular to
the scattering plane.

\subsection{Vector momentum distribution and spin field}

For this work we assume that the proton is knocked-out from the
$1d_{3/2}$ shell of $^{39}$K and the final nucleus $^{38}$Ar is left
in its ground state with total spin $J_B=0$.  In the shell-model
description of this process it was shown in \cite{Ama98a} that the
partial momentum distribution is proportional to the single-particle
distribution of the $1d_{3/2}$ shell (for other values of the final
nuclear spin or for knock-out from inner shells this is no
longer true, although the following results can be generalized to include
these more general cases).  Hence, to simplify the following discussion, we
shall assume that the initial proton is described by a single-particle
wave function with quantum numbers
$(lj)$, and that it is polarized in the direction $\nOmega^*$, {\it i.e.,}
it carries all of the nuclear polarization.

This wave function can be obtained from the corresponding wave function
polarized in the $z$-direction ({\it i.e.,} with magnetic number $m=j$) 
by applying the appropriate rotation operator $R$.
In fact, we begin with a spatial rotation ${\cal R}$ that maps the $z$-axis
onto the polarization direction $\Omega^*$. The rotated
basis, denoted $O^*=(\nne^*_1,\nne^*_2,\nne^*_3)$, is defined  
by
\begin{equation}
\nne^*_i = {\cal R}(\nne_i)
\end{equation}
with the condition 
\begin{equation}
\nne^*_3 = \nOmega^* .
\end{equation}
This last condition does not determine the rotation ${\cal R}$ uniquely, 
since any further rotation around $\nne^*_3$ does not modify the
polarization direction. Of course, the final results cannot 
depend on the particular choice of the vectors $\nne^*_1$,
$\nne^*_2$, and so it is unnecessary to specify them. 

Now we consider a single-particle wave function in momentum space (for
the moment we do not consider the spin) that before the rotation is 
$\psi(\np)=\psi(p_1,p_2,p_3)$, where $p_i$ are the coordinates of
$\np$ in the standard basis $O$. After the rotation the wave function
becomes
\begin{equation}
(R\Psi)(\np)=\psi({\cal R}^{-1}(\np)).
\end{equation}
Here we follow the convention of using the calligraphic symbol ${\cal
R}$ for the rotation in the 3-dimensional space, and the symbol $R$
for the representation of the rotation acting in the Hilbert space of
wave functions. Now we note that the $i$-th component of ${\cal
R}^{-1}(\np)$ in the basis $O$ is
\begin{equation}
[{\cal R}^{-1}(\np)]_i = {\cal R}^{-1}(\np)\cdot \nne_i
= \np\cdot{\cal R}(\nne_i)
= \np\cdot\nne^*_i
\equiv p^*_i ,
\end{equation} 
where $p^*_i$ is the $i$-th component of $\np$ in the rotated basis
$O^*$. Then we have the result
\begin{equation}
(R\psi)(\np) = \psi(p^*_1,p^*_2,p^*_3)\equiv \psi(\np^*) ,
\end{equation}
which simply means that when computing the value of the rotated wave
function for momentum 
$\np=(p_1,p_2,p_3)$ we first compute the components $p^*_i$ of $\np$
in the rotated basis, and evaluate the former wave function 
for momentum $\np^*=(p^*_1,p^*_2,p^*_3)$.

Next we consider the case of a wave function for a particle with spin
$\frac12$
\begin{equation}
\psi_0(\np)=f_1(\np)|\uparrow\rangle+f_2(\np)|\downarrow\rangle .
\end{equation}
Under a rotation the functions $f_i(\np)$ and the spin vectors 
$|\uparrow\rangle$, $|\downarrow\rangle$ also rotate, and
therefore the rotated state $\psi\equiv R\psi_0$ can be written as
\begin{equation}
\psi(\np)=f_1(\np^*)|\uparrow *\rangle+f_2(\np^*)|\downarrow*\rangle ,
\end{equation}
where we have defined the rotated spin basis 
\begin{equation}
|\uparrow *\rangle=
R|\uparrow \rangle,
\kern 2cm
|\downarrow *\rangle=
R|\downarrow \rangle.
\end{equation}
Then the polarized momentum distribution spin matrix is given by
\begin{eqnarray}
n(\np) &=& \psi(\np)\psi(\np)^{\dagger}
\nonumber\\
&=& 
|f_1(\np)|^2 |\uparrow*\rangle\langle\uparrow*|
+|f_2(\np)|^2 |\downarrow*\rangle\langle\downarrow*|
\nonumber\\
&&
\mbox{}
+f_1(\np^*) f_2^*(\np^*) |\uparrow*\rangle\langle\downarrow*|
+f_1^*(\np^*) f_2(\np^*) |\downarrow*\rangle\langle\uparrow*|.
\end{eqnarray}
Hence the problem is reduced to the computation of the four spin
matrices that appear in the above equation. For the first one we have
\begin{equation}
|\uparrow*\rangle\langle\uparrow*|
=R |\uparrow\rangle\langle\uparrow| R^{\dagger} .
\end{equation}
Now we note that
\begin{equation}
|\uparrow\rangle\langle\uparrow| = \frac{1+\sigma_3}{2}
=\frac{1+\nsigma\cdot\nne_3}{2}
\end{equation}
and perform the rotation taking into account the fact that $\nsigma$ 
is a vector operator satisfying
\begin{equation}
R\nsigma R^{\dagger}
= {\cal R}^{-1}(\nsigma) ,
\end{equation}
{\it i.e.,} $\nsigma$ transforms as a 3-dimensional vector with the inverse
rotation, from which we have
\begin{equation}
R\nsigma\cdot\nne_3 R^{\dagger}= {\cal R}^{-1}(\nsigma)\cdot\nne 
= \nsigma\cdot{\cal R}(\nne_3) = \nsigma\cdot\nne_3^* .
\end{equation}
Therefore, we have the result
\begin{equation}
|\uparrow*\rangle\langle\uparrow*| 
=\frac{1+\nsigma\cdot\nne_3^*}{2}.
\end{equation}
The same procedure can be applied to the three remaining spin
operators, obtaining 
\begin{eqnarray}
|\downarrow*\rangle\langle\downarrow*| 
=\frac{1-\nsigma\cdot\nne_3^*}{2}
\\
|\uparrow*\rangle\langle\downarrow*| 
=\frac{\nne_1^*+i\nne_2^*}{2}\cdot\nsigma
\\
|\downarrow*\rangle\langle\uparrow*| 
=\frac{\nne_1^*-i\nne_2^*}{2}\cdot\nsigma
\end{eqnarray}
and the momentum distribution can be written 
\begin{eqnarray}
n(\np) &=& 
|f_1(\np^*)|^2
\frac{1+\nsigma\cdot\nne_3^*}{2}
+|f_2(\np^*)|^2
\frac{1-\nsigma\cdot\nne_3^*}{2}
\nonumber\\
&&
\mbox{}
+f_1(\np^*)f_2^*(\np^*)
\frac{\nne_1^*+i\nne_2^*}{2}\cdot\nsigma
+f_1^*(\np^*)f_2(\np^*)
\frac{\nne_1^*-i\nne_2^*}{2}\cdot\nsigma
\nonumber\\
&=&
\frac12[M^S(\np)+\nM^V(\np)\cdot\nsigma] ,
\end{eqnarray}
where finally the scalar and vector momentum distributions are given
by
\begin{eqnarray}
M^S(\np) 
&=& 
|f_1(\np^*)|^2
+|f_2(\np^*)|^2
\\
\nM^V(\np) 
&=& 
\left\{|f_1(\np^*)|^2-|f_2(\np^*)|^2\right\}\nne_3^*
+2{\rm Re}[f_1^*(\np^*)f_2(\np^*)]\nne_1^*
\nonumber\\
&& \mbox{}
+2{\rm Im}[f_1^*(\np^*)f_2(\np^*)]\nne_2^* .
\label{e55}
\end{eqnarray}

Now we are able to relate the vector momentum distribution to the
spin distribution or spin density. The expectation value of the spin for
the wave function $\psi$ is 
\begin{eqnarray}
\langle\psi|\nS|\psi\rangle 
&=& 
\sum_{rs}\int d^3 p \,
\psi_r^*(\np) \frac{\nsigma_{rs}}{2} \psi_s(\np)
\nonumber\\
&=& \frac12 \int d^3 p\, {\rm Tr}\, [\nsigma n(\np)]
\nonumber\\
&=& \frac12 \int d^3 p\, {\rm Tr}\, 
\left[\nsigma \frac{M_S+\nM_V\cdot\nsigma}{2} \right]
\nonumber\\
&=& \frac12 \int d^3 p\, \nM_V(\np) .
\end{eqnarray}
Hence the vector field $\frac12 \nM_V(\np)$ represents the spin
density in momentum space. Obviously, the scalar momentum distribution
$M_S(\np)=\psi^{\dagger}(\np)\psi(\np)$ is the particle density in
momentum space. The spin density is related to the particle density in
the following way: we define a vector field by the ratio between
the vector and scalar momentum distributions 
\begin{equation}
\ns(\np) \equiv \frac{\nM_V(\np)}{M_S(\np)} .
\end{equation}
This vector always has unit length, $|\ns(\np)|=1$. In fact, we have
\begin{eqnarray}
|M_V|^ 2 
&=& (|f_1|^2-|f_2|^2)^2
            +(2{\rm Re}\, f_1^*f_1)^2
            +(2{\rm Im}\,f_1^*f_2)^2
\nonumber\\
&=& |f_1|^4+|f_2|^4 -2|f_1|^2|f_2|^2+4|f_1^*f_2|^2
\nonumber\\
&=& (|f_1|^2+|f_2|^2)^2
\nonumber\\
&=& 
(M_S)^2 .
\end{eqnarray}
Then we have proven the result
\begin{equation}
\nM_V(\np) = M_S(\np)\ns(\np) ;
\end{equation}
that is, the vector momentum distribution is equal to the scalar 
momentum distribution times a unitary vector field $\ns(\np)$ which
carries the local direction of the spin.

\subsection{Nucleon orbit and spin for the $d_{3/2}$ shell}

Now we consider the specific case of a particle with angular momenta
$(lj)$ which is polarized in the direction $\nOmega^*$. We write the
wave function before the rotation as
\begin{equation}
\psi_0(\np)
= \frac{1}{i^l}\sum_{\mu M}
\langle \textstyle \frac12\mu lM|jj\rangle
Y_{lM}(\hp)\tilde{R}(p)|\frac12\mu\rangle .
\end{equation}
Under a rotation 
the radial function $\tilde{R}(p)$ is invariant, while
the spherical harmonic becomes $Y_{lM}(\hp^*)=
Y_{lM}(\theta_p^*,\phi_p^*)$. Here we denote by
$\hp^*=(\theta_p^*,\phi_p^*)$ the polar and azimuthal angles of the
vector $\np^*=(p_1^*,p_2^*,p_3^*)$ or, equivalently, the polar and
azimuthal angles of $\np$ referred to the new basis $O^*$, {\it i.e.,} the
angles $(\theta_p^*,\phi_p^*)$ are defined by 
\begin{equation}
\np = 
p\sin\theta_p^*\cos\phi_p^* \,\nne_1^*
+p\sin\theta_p^*\sin\phi_p^* \,\nne_2^*
+p\cos\theta_p^* \,\nne_3^* .
\end{equation}
In particular, $\theta_p^*$ is the angle between $\np$ and
$\nOmega^*$.
The rotated wave function $\psi=R\psi_0$ can then be written as
\begin{equation}
\psi(\np) = 
f_1(\np^*)|\uparrow *\rangle
+f_2(\np^*)|\downarrow *\rangle,
\end{equation}
where
\begin{eqnarray}
f_1(\np^*) &=&
\frac{1}{i^l}
\langle \textstyle \frac12\frac12 l,j-\frac12|jj\rangle
Y_{l,j-\frac12}(\theta_p^*,\phi_p^*)
\tilde{R}(p)
\\
f_2(\np^*) &=&
\frac{1}{i^l}
\langle \textstyle \frac12,-\frac12 l,j+\frac12|jj\rangle
Y_{l,j+\frac12}(\theta_p^*,\phi_p^*)
\tilde{R}(p) .
\end{eqnarray}

An illustrative example of the above formalism 
is the ``stretched'' case $j=l+\frac12$, for instance
the $s_{1/2}$, $p_{3/2}$ or $d_{5/2}$ shells. This situation is
particularly simple because $f_2=0$ and therefore, from Eq.~(\ref{e55}),
the vector momentum distribution has no components in the plane 
$(\nne_1^*,\nne_2^*)$. The scalar momentum distribution is 
\begin{equation}
M_S(\np)=|f_1(\np^*)|^2
=|Y_{ll}(\theta_p^*,\phi_p^*)|^2|\tilde{R}(p)|^2
\end{equation}
and for the spin direction we have simply
\begin{equation}
\ns(\np)=\nne_3^*=\nOmega^* .
\end{equation}
In this case the spin is always aligned along the polarization
direction because spin and orbital angular momentum 
must sum to the maximum value $j=l+\frac12$.

Now we study the case of the $d_{3/2}$ shell which is of interest for
the present work of knockout from $^{39}$K. This is a ``jack-knifed''
situation in which the spin and orbital angular momentum are not
aligned. We have for the two functions $f_i(\np^*)$
\begin{eqnarray}
f_1(\np^*) 
&=& 
\sqrt{\frac{3}{8\pi}}
\sin\theta_p^*\cos\theta_p^*
e^{i\phi_p^*}\tilde{R}(p)
\\ 
f_2(\np^*) 
&=& 
\sqrt{\frac{3}{8\pi}}
\sin^2\theta_p
e^{2i\phi_p^*}\tilde{R}(p).
\end{eqnarray}
>From here we find for the scalar momentum distribution
\begin{equation}\label{scalar-momentum}
M_S=\frac{3}{8\pi}\sin^2\theta_p^*|\tilde{R}(p)|^2
\end{equation}
and for the spin direction field
\begin{equation} \label{e70}
\ns(\np)=
\sin 2\theta_p^*\cos\phi_p^* \,\nne_1^*+
\sin 2\theta_p^*\sin\phi_p^* \,\nne_2^*+
\cos 2\theta_p^* \,\nne_3^* .
\end{equation}
This latter equation has the following geometrical meaning: we note that 
the spin vector $\ns(\np)$ is contained in the plane defined by the
momentum $\np$ and the polarization direction $\nOmega^*$, and that
the angle between $\ns(\np)$ and $\nOmega^*$ is twice the angle between
$\np$ and $\nOmega^*$. This means that $\np$ points into the bisectrix
between $\nOmega^*$ and $\ns(\np)$ and, since both $\nOmega^*$ and
$\ns(\np)$ are unit vectors, we must have $\nOmega^*+\ns(\np)=a\np$,
where $a$ is some function of $\np$. To find the latter we compute the square
\begin{equation}
1=\ns^2 = (a\np-\nOmega^*)^2 = a^2p^2+1-2ap\cos\theta_p^* ,
\end{equation}
from which we obtain $a=(2/p)\cos\theta_p^*$ and therefore
\begin{equation} \label{spin}
\ns(\np) = 2\frac{\nOmega^*\cdot\np}{p^2}\np-\nOmega^* .
\end{equation}
This depends only on $\np$ and $\nOmega^*$ and is
independent of the particular directions chosen for $\nne_1^*$ and
$\nne_2^*$, as expected.

In the semi-classical model, the quantity of interest
is the expectation value of the position for a
nucleon with given missing momentum, defined by \cite{Ama99a}
\begin{equation}
\nr(\np) = 
\frac{\psi^{\dagger}(\np)(i\nabla_p)\psi(\np)}
     {\psi(\np)^{\dagger}\psi(\np)} ,
\end{equation}
where we have used the momentum-space representation of the position
operator $\hat{\nr}=i\nabla_p$. In the case of a particle 
in the $d_{3/2}$ shell polarized in the $\nOmega^*$-direction, 
the expected position was computed in \cite{Ama99a} and can be written
as
\begin{equation} \label{expected-position}
\nr(\np)= -\frac{1+\sin^2\theta^*_p}{p^2\sin^2\theta^*_p}
\nOmega^*\times\np .
\end{equation}

Similar expressions can be written in coordinate space. The 
spatial density is given by
\begin{equation}\label{density}
\rho(\nr)=\frac{3}{8\pi}\sin^2\theta^*_r |R(r)|^2 ,
\end{equation}
where $\theta^*_r$ is now the angle between $\nOmega$ and $\nr$,
while the expectation value of momentum for given position can be
obtained
from the momentum operator in position space $\np=-i\nabla_r$ as
\begin{equation} \label{expected-momentum}
\np(\nr)= \frac{1+\sin^2\theta^*_r}{r^2\sin^2\theta^*_r}
\nOmega\times\nr.
\end{equation}

\subsection{Applications to the HD cross section}

Now we have at hand all of the ingredients that enter in the
semi-classical description of the HD cross section. We first discuss
the in-plane emission case, $\phi=0^{\rm o}$ where nodal symmetry in the HD cross
section was found in the last section. 

In Fig.~11 we show a schematic picture of what is happening for the
case of $(\theta^*,\Delta\phi)=(90^{\rm o},90^{\rm o})$, corresponding to nuclear
polarization pointing to the $-y$ direction. The torus-like region
represents the nucleon orbit corresponding to the $d_{3/2}$ shell,
determined by the nucleon density in coordinate space, 
Eq.~(\ref{density}).  This ``orbit'' represents the spatial region where
it is most probable to find the nucleon. The radius of the orbit is
determined by the maximum of the radial wave function for the
$d_{3/2}$ shell.  Within the semi-classical model, the nucleon 
moves circularly around the rotation axis represented by the
polarization vector $\nOmega^*$, as follows from
Eq.~(\ref{expected-momentum}). In the case of Fig.~11 the sense of
circular movement is clockwise in the $zx$ plane.  Hence the
orientation and sense of the nucleon orbit is fixed by the nuclear
polarization.  From the setup of the reaction kinematics we also know the
missing momentum $\np=\np'-\nq$ of the proton, which in the figure is
almost perpendicular to $\nq$, corresponding to quasielastic
kinematics as in the results of the previous section.  This in fact
determines the most probable position of the proton before the
interaction, as follows from Eq.~(\ref{expected-position}) (in Fig.~11
the proton is shown schematically as a white ball).  It is located in
the lower part of the orbit with respect to the exit direction, as
determined by the final momentum $\np'$, and so the FSI in this case
are expected to be large, since it has to cross the entire nucleus
before exiting, as was shown for the helicity-independent cross
section in \cite{Ama99a}.

\begin{figure}[hptb]
\begin{center}
\leavevmode
\def\epsfsize#1#2{0.7#1}
\epsfbox{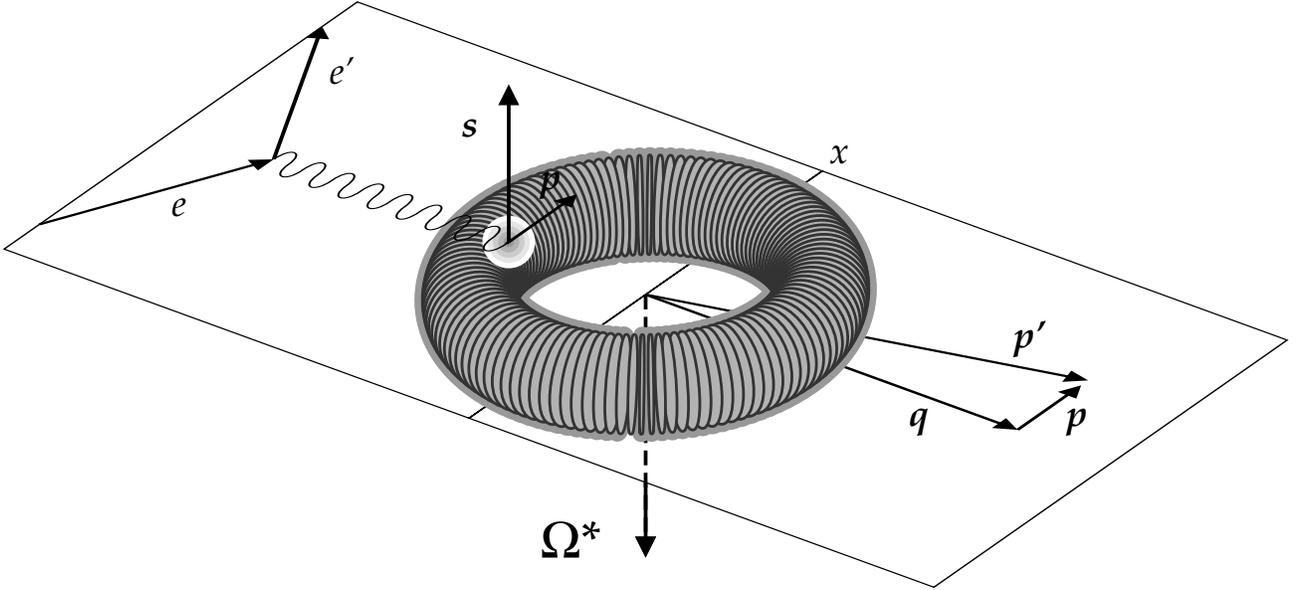}
\end{center}
\caption{
Nucleon orbit for in-plane emission and nodal polarization 
$(\theta^*,\Delta\phi)=(90^{\rm o},90^{\rm o})$. The expected value of
the proton's location for the given missing momentum 
is represented by the white ball. In this case the nucleon spin is
opposite to the polarization vector $\nOmega^*$. The $z$-axis lies in
the direction of {\bf q}.
}
\end{figure}

Now in the case of the HD cross section a new element comes into
play, namely the spin vector, Eq.~(\ref{spin}), which for
the kinematics of Fig.~11 is given by $\ns=-\nOmega$, since $\np$ is
perpendicular to $\nOmega$.

In PWIA the HD response functions in Eq.~(\ref{RK'}) are 
proportional to the scalar product 
between $\ns(\np)$ and a single-nucleon vector,
Eqs.~(\ref{wt'},\ref{wtl'}). For in-plane emission, $\phi=0^{\rm o}$, 
the pieces containing $\sin\phi$ do not contribute and hence the 
single-nucleon vectors are
\begin{eqnarray}
\nw_V^{T'}  
& = &  2J_cJ_m\delta\,\nne_1-2J_m^2\nne_3 
\label{wt'phi=0}\\
\nw_V^{TL'} 
& = & 2\sqrt{2}\rho_cJ_m\nne_1-2\sqrt{2}\rho_{so}J_m\delta\,\nne_3 .
\label{wtl'phi=0}
\end{eqnarray}
An important result that follows from these equations is that {\em both
single-nucleon vectors lie in the scattering plane}. Hence, the HD
cross section is determined only by the
projection of the nucleon spin in the scattering plane. Since in
Fig.~11 the spin is perpendicular to this plane, we have $\Delta=0$ 
for the nodal polarization $(90^{\rm o},90^{\rm o})$ and the same happens for the
opposite polarization $(270^{\rm o},270^{\rm o})$, where the spin is inverted.

\begin{figure}[hptb]
\begin{center}
\leavevmode
\def\epsfsize#1#2{0.7#1}
\epsfbox{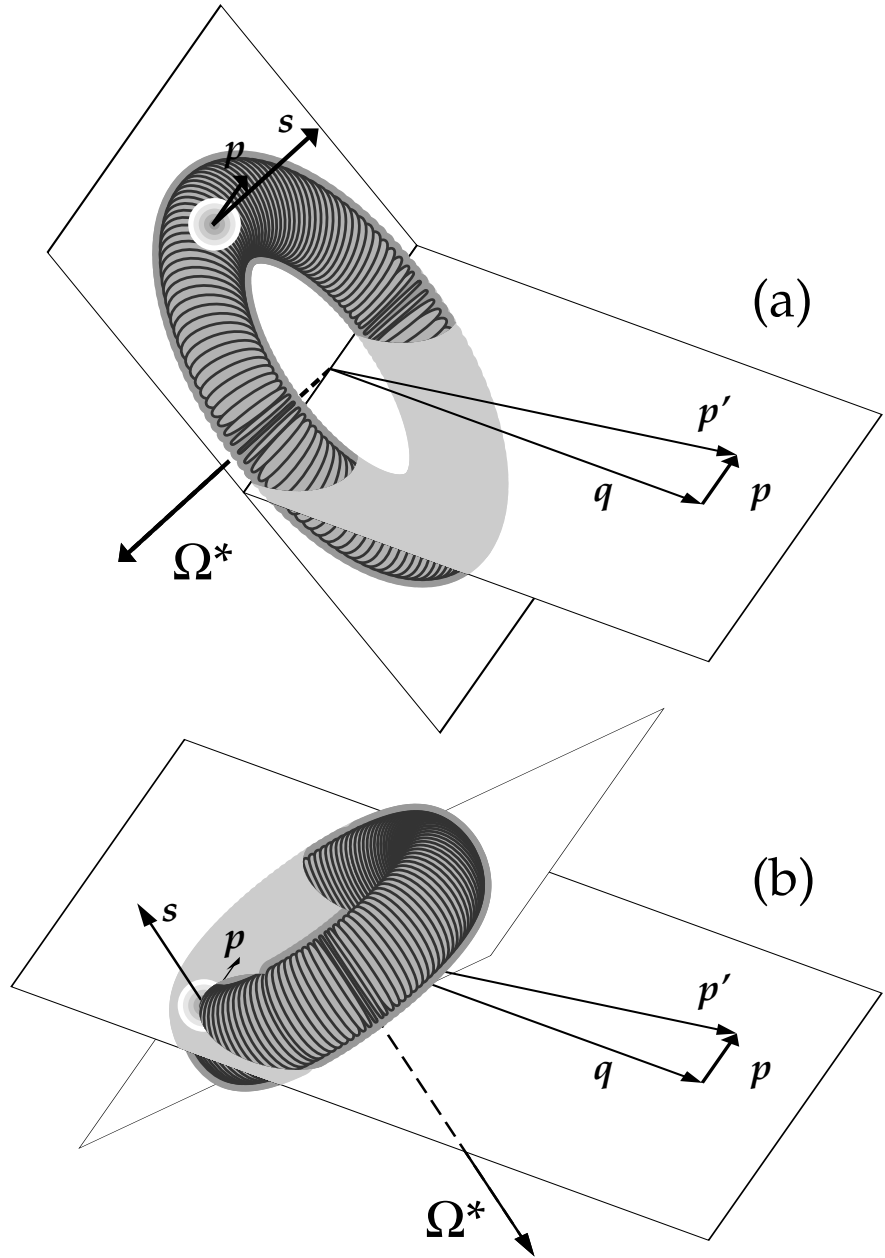}
\end{center}
\caption{
Nucleon orbit for two nuclear polarizations that are symmetrical with respect
to the nodal polarization, obtained by $\pm 45^{\rm o}$ 
rotation of the orbit of
Fig.~11 with respect to the $x$-axis. Also shown are the 
nucleon  location and spin for the given missing momentum. 
}
\end{figure}

The origin of the nodal symmetry must come from pairs of
polarizations with different spins but opposite projections on the
scattering plane. An example is show in Fig.~12, corresponding to $\pm
45^{\rm o}$ rotations of the nucleon orbit around the $x$-axis.
Fig.~12(a) corresponds to the polarization
$(\theta^*,\Delta\phi)=(135^{\rm o},90^{\rm o})$, while (b) stands for
$(45^{\rm o},90^{\rm o})$.  These
polarizations correspond to a pair of panels in Fig.~1.  In Fig.~12 we
see that the spin vectors for (a) and (b) point in different
directions, although their projections on the scattering plane are 
equal and opposite. We also see that 
the proton is located at the same point in the respective orbits,
hence with equal values of the scalar momentum distribution.
Therefore the corresponding HD cross sections are expected to be equal
and opposite in PWIA, as was found for the results shown in Fig.~1.  Now the
question of why the symmetry persists even in presence of FSI can also
be answered using the present example. In fact, from Fig.~12 we see that
the proton in (a) and (b) is located in symmetrical positions with
respect to the $x$-$z$ plane, above and below the plane,
respectively. In the two cases the proton exits the nucleus with the same
momentum $\np'$ from positions that are symmetrical with respect to
it, traversing equivalent paths across the nucleus and hence
experiencing the same interaction (absorption) due to the imaginary
part of the central optical potential. Since in both cases the initial
impact parameter is also the same, the interaction due to the real
part of the central optical potential is also the same.  

It remains to be shown that the effect of the spin-orbit part of the
potential is also the same in the two cases. This is not obvious {\em
a priori} because the proton spins in (a) and (b) are
different. However, the scalar product $\nl\cdot \ns$ is what enters
in the spin-orbit interaction, and remarkably this scalar product is
also the same in both cases. In fact, in the semi-classical model the
proton exits the nucleus with angular momentum $\nl=\nr\times\np'$ and
initially with spin $\ns$. Hence the spin-orbit interaction is 
\begin{equation}
V_{ls}(r)(\nr\times\np')\cdot\ns = -V_{ls}(r)(\nr\times\ns)\cdot\np' .
\end{equation}
In Fig.~12 we see that the value of the vector $\nr\times\ns$ 
is the same both in cases (a) and (b), since so is the angle between 
the vectors $\nr$ and $\ns$, 
and the sense of the vector product is in the $x$-direction. 

Therefore we have proven within the semi-classical model 
that the FSI effects are the same for the 
two orbit orientations displayed  in Fig.~12. These effects are manifested 
in a modification of the PWIA cross section (both the
helicity-dependent and -independent pieces), producing a change of the
strength and redistribution of the shape. Since the HD cross section in
PWIA is opposite for the two polarizations of Fig.~12 and the secondary
effect of FSI is the same in the two cases, the HD cross section
that results from  including the FSI in DWIA is still equal and opposite  in
both cases.

\begin{figure}[hptb]
\begin{center}
\leavevmode
\def\epsfsize#1#2{0.8#1}
\epsfbox[20 320 520 700]{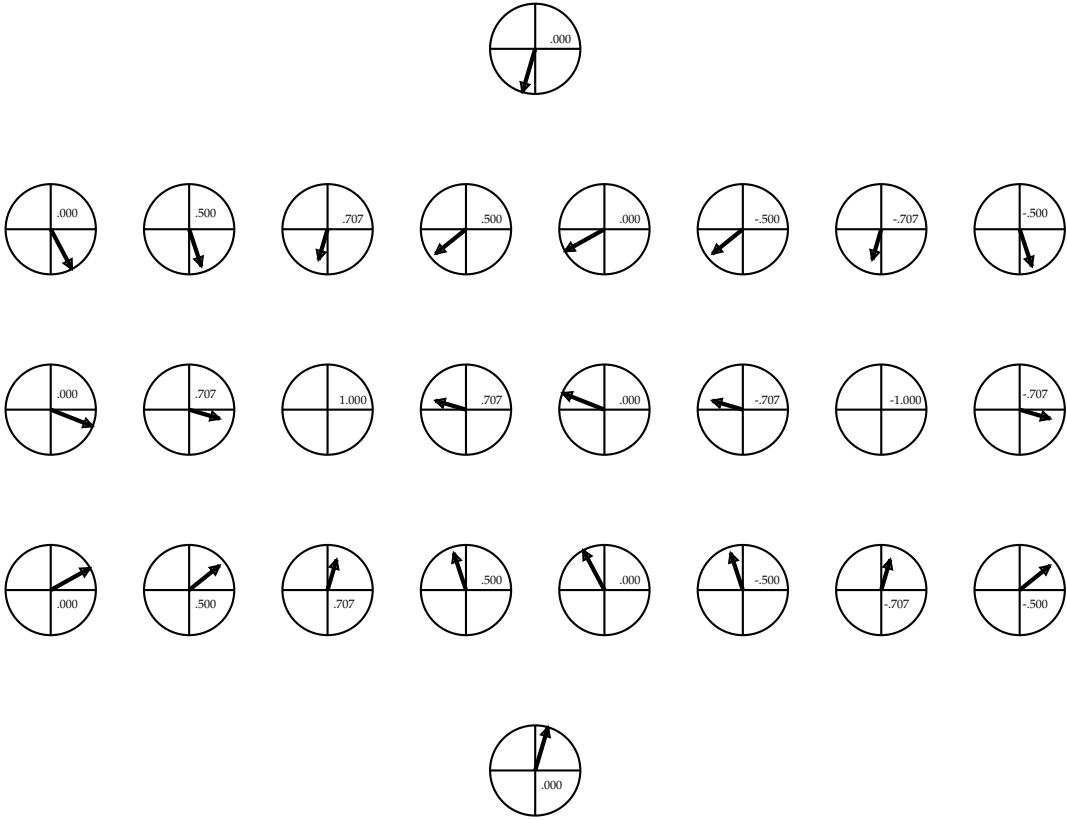}
\end{center}
\caption{ Spin direction of the proton computed using
Eq.~(\protect\ref{spin}) for each of the polarizations of Figs.~1
and 2, for $\phi=0^{\rm o}$, and for missing momentum $p=180$ MeV/c.
With an arrow we show the projection of the spin on the $x$-$z$ plane.
The inserted numbers indicate the value of the $y$ component in
each case. }
\end{figure}

The existence of the nodal symmetry and its persistence in the presence
of FSI can be explained in a similar geometrical way for all of the
polarizations shown in Figs.~1 and 2. This can be done by noting that for
each pair of nodal polarizations the respective orbits can be obtained
by performing a rotation of the nodal orbit of Fig.~11 by angles $\pm
\tilde{\theta}$ with respect to an axis contained in the $x$-$z$
plane. For instance, in the case of Fig.~12, the $\pm 45^{\rm o}$ 
rotations were performed with respect to the $x$-axis. 
For a fixed value of the missing momentum, the value of
$\sin\theta^*_p$
(remember that  $\theta^*_p$ is the angle between $\np$ and $\nOmega^*$)
is the same for the two orbits rotated by angles $\pm\tilde\theta$,
as is shown in the Appendix. Therefore, 
the value of the  scalar momentum
distribution given by Eq.~(\ref{scalar-momentum}) 
is the same for both orbits, since its angular dependence is 
$\sin^2\theta^*_p$. 
As a consequence, the helicity-independent response functions and
cross section are also equal for this pair of polarizations. 
In the case of the HD responses there is an additional dependence on 
the nucleon spin via the scalar product $\nw^K_V\cdot\ns(\np)$. 
In Fig.~13 we show the spin vector for each one of the polarizations
of Figs.~1 and 2, computed using Eq.~(\ref{spin}), 
for a value of the missing momentum fixed to $p=180$ MeV/c.
In the figure we can see that the projection of the spin on the 
scattering plane is opposite for each pair of polarizations lying symmetrical
with respect to the nodal one, for which $\ns$ is perpendicular to the
plane. 
Since the single-nucleon vectors $\nw^K_V$ are contained in this
plane, we find opposite values of $\nw^K_V\cdot\ns$ for each pair of
symmetrical polarizations. This proves that the HD responses are odd
with respect to the nodal symmetry in PWIA. 
An analytical proof of this result is provided in the Appendix.

In arguing that the nodal symmetry persists in presence of the
FSI in Fig.~14 we show a plot of the nucleon position within the
nucleus and spin vectors for the polarizations of Figs.~1 and 2. 
The nucleon position has been computed using Eq.~(\ref{expected-position})
for $p=180$ MeV/c. To help in the visualization, in each plot of Fig.~14 
we display the  exit plane spanned by the vectors $\nr$ and $\np'$,
in a coordinate system with the $y$-axis pointing in the $\np'$ direction.
In the figure we can see that the initial position of the 
nucleon is the same for each pair of polarizations lying symmetrical with
respect to the nodal point, which implies the same FSI due to the
central part of the optical potential for these polarizations. 
This result is proven analytically in the Appendix.

\begin{figure}[hptb]
\begin{center}
\leavevmode
\def\epsfsize#1#2{0.8#1}
\epsfbox[40 310 540 710]{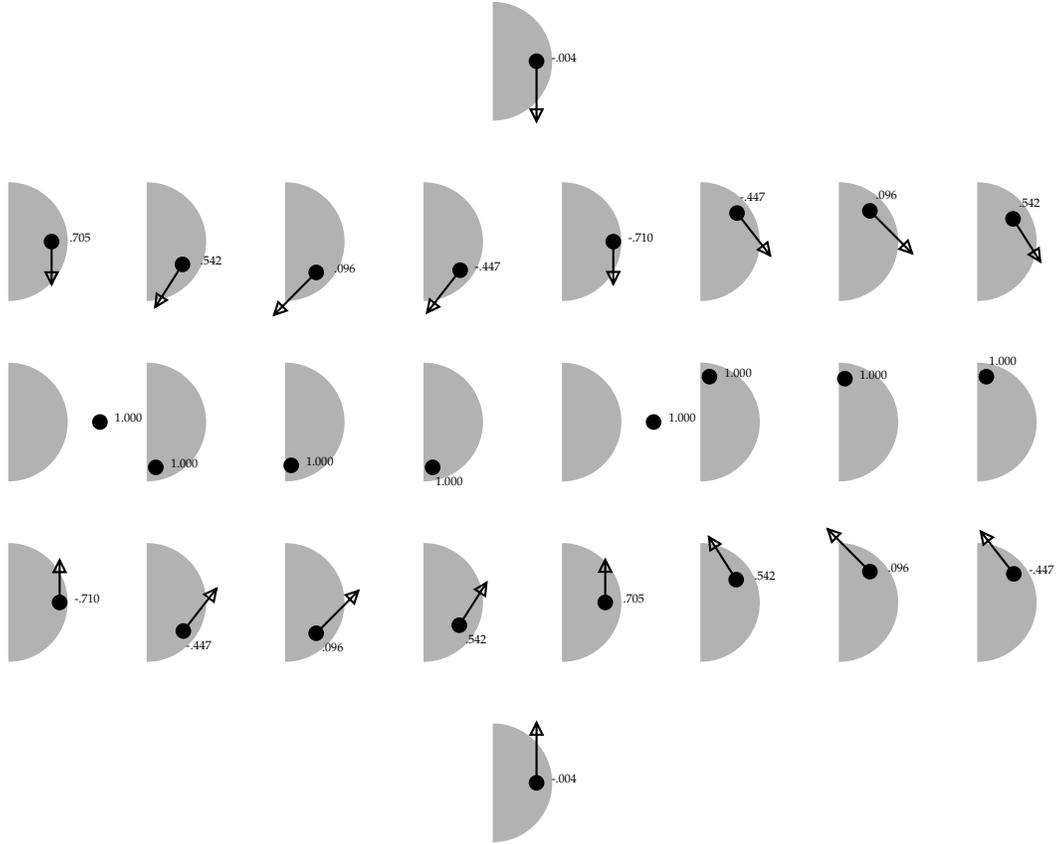}
\end{center}
\caption{ Expected value of position computed using
Eq.~(\protect\ref{expected-position}) for each of the
polarizations of Figs.~1 and 2, for $\phi=0^{\rm o}$ and for missing
momentum $p=180$ MeV/c. The gray semi-circles represent half a slice
of the nuclear interior.  The coordinate system in the figure is
chosen in each case to be the exit plane spanned by the vectors $\nr$
and $\np'$, which is pointing in the $y$-direction in all cases.  The
arrows represent the spin projection on this plane, while with the
inserted numbers we indicate the perpendicular component of the
spin. }
\end{figure}

Moreover, the effect of the spin-orbit FSI can also be
inferred from the figure, where we show the projection of the spin on
the exit plane.  Since the final angular momentum $\nr\times\np'$ is
perpendicular to the exit plane, the spin-orbit coupling $\nl\cdot\ns$
is determined only by the perpendicular component of the spin
(indicated with the inserted numbers in Fig.~14), and where again we can see
that these values are the same for pairs of polarizations that are symmetrical
with respect to the nodal point. This implies the same spin-orbit
interaction for these polarizations, and is also proven analytically in the
Appendix.

These arguments show that the origin of the nodal symmetry in the HD
response functions and $\Delta$ in the presence of FSI 
can be explained in the semi-classical model for $\phi=0^{\rm o}$. 
Next we discuss the case $\phi=90^{\rm o}$.  
We have seen in the previous section that in this case 
the nodal symmetry is broken for the HD cross section $\Delta$ because 
the $T'$ and $TL'$ responses present odd and even nodal symmetry
respectively. The reason for this different behavior of the $T'$ and
$TL'$ response functions comes from the single-nucleon vectors,
Eqs.~(\ref{wt'},\ref{wtl'}), which
for $\phi=90^{\rm o}$ can be written as
\begin{eqnarray}
\nw_V^{T'} 
&=&
2J_c J_m \delta\,\nne_2-2J_m^2\nne_3
\label{wt'90}\\
\nw_V^{TL'} 
&=& 2\sqrt{2}(\rho_c J_m -J_c\rho_{so}\delta^2)\nne_1 .
\label{wtl'90}
\end{eqnarray}
Note the essential difference between the former case, $\phi=0^{\rm o}$, where
both single-nucleon vectors were contained in the scattering plane,
and the present case, where the $T'$ and $TL'$ vectors are
perpendicular, with the latter pointing to the $x$-direction. 

\begin{figure}[hptb]
\begin{center}
\leavevmode
\def\epsfsize#1#2{0.7#1}
\epsfbox{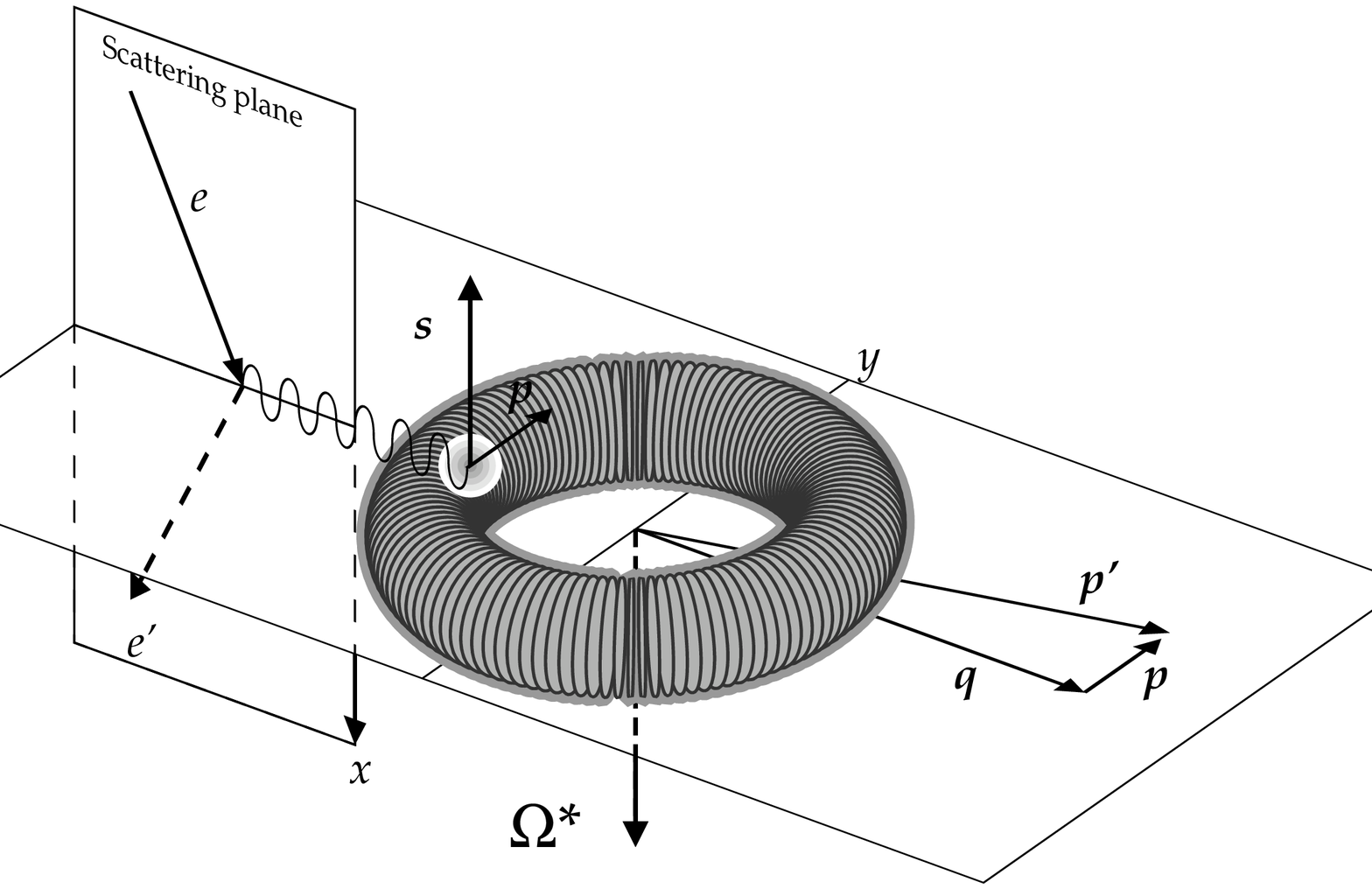}
\end{center}
\caption{ Nucleon orbit for out-of-plane emission, $\phi=90^{\rm o}$,
and for the nodal polarization 
$(\theta^*,\Delta\phi)=(90^{\rm o},90^{\rm o})$.  
Note that the only geometrical difference from Fig.~11 is that 
in this case the scattering plane is perpendicular to the reaction
plane.
}
\end{figure}

The differences from the case $\phi=0^{\rm o}$ can be better appreciated by
looking at Fig.~15 where it is clear that now the scattering plane ($xz$)
is perpendicular to the reaction plane ($yz$). The nucleon orbit
shown in Fig.~15 corresponds to the new nodal polarization
$(\theta^*,\Delta\phi)=(90^{\rm o},90^{\rm o})$.  Note that in this case
$\Delta\phi=90^{\rm o}-\phi^*$, so the nodal polarization corresponds
to $\phi^*=0^{\rm o}$, {\it i.e.,} 
$\nOmega^*=\nne_1$, and the nucleon orbit lies in
the reaction plane. Comparing Fig.~15 with Fig.~11 we note that the
only difference between them lies in the scattering plane spanned 
by  the electrons, which in one case coincides with the reaction plane
and is perpendicular in the other. However, {\em the mechanisms 
underlying the reaction that have to do with the proton orbit and FSI,
viz., those that occur in the reaction plane, are identical in both cases.} 

In fact, we can easily transform the $\phi=90^{\rm o}$ case into
the $\phi=0^{\rm o}$ one by just making a change of coordinates
\begin{equation}
x' = y,   \kern 2cm  y'=-x, \kern 2cm z'=z
\end{equation}
under which the reaction plane is now the $x'z'$ plane.
Therefore all of the details
studied in the case $\phi=0^{\rm o}$ concerning the nucleon orbit, position,
spin and FSI effects are also valid  in the present case. 
In particular, Fig.~13 represents the components of the spin 
in the new reaction plane, corresponding to the polarizations of 
Figs.~3 and 4, while the proton positions and components of the spin in the exit
plane shown in Fig.~14 are also valid in this case. 

The only difference with the former case is hence in the single-nucleon
vectors of Eqs.~(\ref{wt'90},\ref{wtl'90}), 
which in the present case written in the new coordinates are given by
\begin{eqnarray}
\nw_V^{T'} 
&=&
2J_c J_m \delta\,\nne'_1-2J_m^2\nne'_3
\\
\nw_V^{TL'} 
&=& -2\sqrt{2}(\rho_c J_m -J_c\rho_{so}\delta^2)\nne'_2 .
\end{eqnarray}
Hence the $T'$ vector is  in the $x'z'$ plane, 
while in the former case it was contained in the $xz$ plane. Therefore
the behavior of the $T'$ response function is exactly the same 
as in the $\phi=0^{\rm o}$ case. In particular, it is odd under the nodal
symmetry. This was expected, because it is known that 
this response is only a function of $\Delta\phi$, and this is a
consequence of the fact that the single-nucleon vector $\nw^{T'}_V$ is
always contained in the reaction plane.

The difference hence lies in the $TL'$ response, since now the 
single-nucleon vector $\nw_V^{TL'}$ is perpendicular to the reaction
plane, while in the former case it was contained in that plane. 
For this reason, the behavior of the $TL'$ response in PWIA now depends
on the spin component {\em perpendicular} to the reaction plane, 
which is shown by the inserted numbers in Fig.~13. Therein we 
can see that this spin component is the same 
for pairs of polarizations that are symmetrical with respect to the nodal point.
Therefore in this case there is no change of sign of the $TL'$
response under the nodal symmetry, from which it follows 
that this response is even under that symmetry, as was seen in the
results of Figs.~9 and 10. 

These arguments help to clarify the fact that there are actually 
{\em two} reasons why the fifth response function arising
for $\phi=90^{\rm o}$ is different
from zero for unpolarized nuclei when integrating over all 
orientations of the nucleon orbit: the first, well-known reason is that 
the nucleon's path through the nucleus is different from one polarization
to the opposite and hence undergoes different FSI. The second reason
is that this response involves the component of the nucleon spin
perpendicular to the reaction plane --- {\it i.e.,} in the $x$
direction ---
and this component is the same for pairs of polarizations that are symmetrical
with respect to the nodal one --- corresponding to the nucleon orbit
being contained in the reaction plane.

\section{Conclusions}

In this work we have studied the helicity-dependent part of the cross
section for $(e,e'p)$ reactions with polarized beam and target.  
Using the shell model to describe the nuclear structure involved, we
have presented results of a DWIA calculation for proton knock-out from
the $d_{3/2}$ shell of polarized $^{39}$K under 
quasielastic kinematics for the full range of nuclear polarization
directions. Such an analysis, performed as a function of the polarization
angles, has allowed us to discover a new symmetry 
in the HD cross section that arises for {\em in-plane} emission.
This entails a change of sign for pairs
of polarization vectors that are opposite with respect to the {\em nodal}
polarization ({\it i.e.,} perpendicular to the reaction plane). It is as a
consequence of this nodal symmetry that the fifth response function is
zero for unpolarized nuclei in these conditions even in presence of
FSI.

Our results also show that the nodal symmetry is broken for
{\em out-of-plane} emission. However, we have found that the symmetry is
still present in the three separate response functions $W^{T'}$,
$\widetilde{W}^{TL'}$ and $W^{TL'}$, but is not of a single type:
while the first two responses are odd under the nodal symmetry, the
latter is even, and a linear combination of the three responses
therefore breaks the symmetry in the cross section.

The next goal of the paper has been to explain the origin of the nodal
symmetry within a semi-classical model of the reaction based on the
PWIA and on the concept of a nucleon's orbit with attendant expectation values of
position and spin, and in which the exiting nucleon's path through the
nucleus determines the strength
of the FSI. We have shown that for pairs of nodal-symmetrical
polarizations the nucleon positions and initial impact parameters are
equivalent and with equal values of the spin-orbit coupling
$\nl\cdot\ns$, explaining why the FSI effects are the same for these
polarizations.  The FSI modify the PWIA response functions, which
are determined by the scalar product $\nw^K_V\cdot\ns$, where $\ns$ is
the nucleon spin and $\nw^K_V$ is the single-nucleon vector.  We have
shown that the component of the spin in the reaction plane changes
sign under the nodal symmetry, while the component perpendicular to
this plane is invariant under it. As a consequence, since the $T'$
single-nucleon vector is always contained in the reaction plane, the
$T'$ response is always odd under the nodal symmetry.  On the other
hand, for in-plane emission, the $TL'$ single-nucleon vector is also
contained in the reaction plane, so the $TL'$ response is also odd in these
conditions. However, in the case $\phi=90^{\rm o}$ the $TL'$
single-nucleon vector is {\em perpendicular} to the reaction plane,
and so the corresponding $TL'$ response is {\em even} under the nodal
symmetry. 

As a consequence of the above, the fifth response function for 
scattering of polarized electrons from unpolarized nuclei
(obtained as an average over orbit orientations) is not zero, and
furthermore, in addition to requiring FSI to occur, it probes 
the spin component of the nucleon perpendicular to the reaction plane.

We would like to emphasize that, although the existence of the nodal
symmetry has been found here numerically within the context of a
particular DWIA model and proven analytically only in the
semi-classical model, there is clear evidence that this is a general,
model-independent property of the cross section. This is supported by
the fact that the fifth response function is known theoretically to be
zero for in-plane emission, so the cancellation between pairs of
polarizations must be exact.  To provide an analytical,
model-independent proof of this theorem from general principles
appears not to be a trivial task. However, the results of the present
study are clear motivation to find a general symmetry-based proof
starting from the formalism of \cite{Ama98a,Ras89}.

The success of the present model in explaining the properties of the
HD cross section where the proton spin plays a relevant role opens the
possibility of using this type of reaction to study not only the
3-dimensional momentum distribution --- accessible using polarized
nuclei and unpolarized electrons --- but also the nucleon spin
distribution in nuclei --- using in addition polarized electrons.  The
knowledge gained in the understanding of $(e,e'p)$ spin observables
through this kind of study, where many of the properties of the cross
section, including FSI can be characterized in a clear geometrical
picture of the reaction, can be of great help in the design of future
experiments with polarized nuclei and electrons.  As seen through the
examples presented in this paper, this could play an important role in
advancing our knowledge of nuclear dynamics.

\section*{Appendix: 
Proof of the nodal symmetry in the semi-classical model}

We consider the in-plane emission case, $\phi=0^{\rm o}$, with a fixed
value of the missing momentum $\np$.  Remember that in this case the
single-nucleon vectors $\nw^{T'}_V$ and $\nw^{TL'}_V$ are also
contained in the scattering plane, as they are given by
Eqs.~(\ref{wt'phi=0},\ref{wtl'phi=0}). 
We take as reference the nodal polarization corresponding to 
$\theta^*=90^{\rm o}$, $\phi^*=-90^{\rm o}$, {\it i.e.,} $\nOmega^*=-\nne_2$.
In this case the nucleon orbit lies in the scattering plane as
displayed in Fig.~11. 
If we rotate the nodal orbit by an angle $\tilde{\theta}$ around 
an axis contained in the scattering plane then

i) the helicity-independent cross section $\Sigma$ is the same for 
rotation angles $\pm\tilde\theta$, and

ii) the HD cross section $\Delta$ has opposite signs for
$\pm\tilde\theta$. 

To prove this, let us consider two polarization vectors, 
$\nOmega_1$ and $\nOmega_2$, that are
symmetrical with respect to the nodal polarization
\begin{eqnarray}
\nOmega_1 &=& \sin\tilde\theta\,\nOmega_s-\cos\tilde\theta\,\nne_2
\\
\nOmega_2 &=& -\sin\tilde\theta\,\nOmega_s-\cos\tilde\theta\,\nne_2 ,
\end{eqnarray}
where $\nOmega_s$ is a unitary vector contained in the scattering plane,
{\it i.e.,} 
\begin{equation}
\nOmega_s= \cos\tilde\phi\,\nne_1+\sin\tilde\phi\,\nne_3 .
\end{equation}
Then the following results emerge:

\begin{enumerate}

\item  The projections of $\nOmega_1$ and $\nOmega_2$ over the missing
momentum $\np$ have opposite sign, since $\np$ lies in the scattering
plane, {\it i.e.,} 
\begin{equation}
\nOmega_1\cdot\np = -\nOmega_2\cdot\np =
\sin\tilde\theta\,\nOmega_s\cdot\np.
\end{equation}
Hence the angles $\theta^*_{p1}$ and $\theta^*_{p2}$ between $\np$ and
the polarization vectors $\nOmega_1$ and $\nOmega_2$ satisfy
\begin{eqnarray}
\cos\theta^*_{p1}&=& -\cos\theta^*_{p2}\\
\sin\theta^*_{p1}&=&  \sin\theta^*_{p2} .
\end{eqnarray}

As a consequence, since the scalar momentum distribution for the
$d_{3/2}$ shell is
$M_s=\frac{3}{8\pi}\sin^2\theta^*_p|\tilde{R}(p)|^2$, 
and the value of $\sin\theta^*_{pi}$ is the same for these two
polarizations, the helicity-independent response functions 
in Eq.~(\ref{RK}) are the same for these two polarizations in PWIA, and
hence so are the cross sections $\Sigma$ in PWIA.

\item Let $\ns_1$ and $\ns_2$ be the spin vectors for these two
orbits,  given by Eq.~(\ref{spin}). Then this pair of  spins
has {\em opposite} components on the scattering plane 
and {\em equal} components on the $y$-axis. 
In fact we have, from Eq.~(\ref{spin}),
\begin{eqnarray}
\ns_1 &=& \frac{2\nOmega_1\cdot\np}{p^2}\np-\nOmega_1
= \frac{2\sin\tilde\theta\,\nOmega_s\cdot\np}{p^2}\np
  -\sin\tilde\theta\,\nOmega_s+\cos\tilde\theta\,\nne_2
\\
\ns_2 &=& \frac{2\nOmega_2\cdot\np}{p^2}\np-\nOmega_2
= -\frac{2\sin\tilde\theta\,\nOmega_s\cdot\np}{p^2}\np
  +\sin\tilde\theta\,\nOmega_s+\cos\tilde\theta\,\nne_2 ;
\end{eqnarray}
hence the spin components on the scattering plane, $(\ns_i)_s$, are
opposite and satisfy
\begin{equation}
(\ns_1)_s=-(\ns_2)_s 
= \sin\tilde\theta 
  \left(  \frac{2\nOmega_s\cdot\np}{p^2}\np-\nOmega_s \right) ,
\end{equation}
while the $y$-components are equal and given by
\begin{equation}
(\ns_1)_y= (\ns_2)_y = \cos\tilde\theta\,\nne_2 .
\end{equation}
Since the single-nucleon vectors in Eqs.~(\ref{wt'phi=0},\ref{wtl'phi=0}) are
in the scattering plane, their scalar products with the spin
vectors $\ns_i$ are opposite, as are also the HD response functions
$T'$ and $TL'$ in PWIA; {\it i.e.,} they are odd with respect to the nodal
symmetry. The same result holds for the HD cross section $\Delta$.

\item Let $\nr_1$ and $\nr_2$ the expected values of position of the
nucleon for the two polarizations $\nOmega_1$, $\nOmega_2$. Then
$\nr_1$ and $\nr_2$ have {\em opposite} projections on the scattering plane
and {\em equal} projections on the $y$-axis.
In fact, from Eq.~(\ref{expected-position}) we have 
\begin{eqnarray}
\nr_i= -\frac{1+\sin^2\theta^*_{pi}}{p^2\sin^2\theta^*_{pi}}
\nOmega_i\times\np .
\end{eqnarray}
By result 1, $\sin\theta^*_{p1}=\sin\theta^*_{p2}$ and therefore the 
factor in front of $\nOmega_i\times\np$ is the same for $\nr_1$ and
$\nr_2$. Furthermore, we have for the vector products
\begin{eqnarray}
\nOmega_1\times\np
& = & \sin\tilde\theta\,\nOmega_s\times\np
     -\cos\tilde\theta\,\nne_2\times\np
\\
\nOmega_2\times\np
& = & -\sin\tilde\theta\,\nOmega_s\times\np
     -\cos\tilde\theta\,\nne_2\times\np .
\end{eqnarray}
We compute separately the two vector products above. From the first one
we obtain the projection on the $y$-axis
\begin{eqnarray}
\nOmega_s\times\np
& = & p(\cos\tilde\phi\,\nne_1+\sin\tilde\phi\,\nne_3)
      \times(\sin\theta\,\nne_1+\cos\theta\,\nne_3)
\nonumber\\
&=& p(\sin\tilde\phi\sin\theta-\cos\tilde\phi\cos\theta)\nne_2 ,
\end{eqnarray}
while for the second one we obtain the projection on the scattering plane
\begin{equation}
\nne_2\times\np = p\nne_2\times(\sin\theta\,\nne_1+\cos\theta\,\nne_3)
=p(\cos\theta\,\nne_1-\sin\theta\,\nne_3) .
\end{equation}
Therefore we obtain for the position vectors
\begin{eqnarray}
\nr_1
&=&
\frac{1+\sin^2\theta^*_{p1}}{p\sin^2\theta^*_{p1}}
\left[-\sin\tilde\theta
       (\sin\tilde\phi\sin\theta-\cos\tilde\phi\cos\theta)\nne_2
       -\cos\tilde\theta(\cos\theta\,\nne_1-\sin\theta\,\nne_3)
\right]
\nonumber\\
\\
\nr_2
&=&
\frac{1+\sin^2\theta^*_{p1}}{p\sin^2\theta^*_{p1}}
\left[\sin\tilde\theta
       (\sin\tilde\phi\sin\theta-\cos\tilde\phi\cos\theta)\nne_2
     +\cos\tilde\theta(\cos\theta\,\nne_1-\sin\theta\,\nne_3)
\right] .
\nonumber\\
\end{eqnarray}
From these equations we obtain the desired result.

\item The projections of $\nr_1$ and $\nr_2$ over the exit vector
$\np'$ are the same, and the exit impact parameters $b_1$ and $b_2$
are equal. In fact,  the first result 
follows from point 3 above, since $\np'$ is in the
scattering plane and both vectors $\nr_1$ an $\nr_2$ have the same
component in this plane. 
On the other hand, the impact parameter  is defined with respect to
the exit direction $\np'$ as the modulus of the perpendicular
component of $\nr$ with respect to $\np'$, {\it i.e.,} 
\begin{equation}
\nb = \nr-\frac{\nr\cdot\np'}{p'{}^2}\np' .
\end{equation}
Since $\nr_1\cdot\np'=\nr_2\cdot\np'$ and, from point 3,
$|\nr_1|=|\nr_2|$, the components of $\nr_1$ perpendicular to $\np'$
are equal, {\it i.e.,} $b_1=b_2$. 

As a direct consequence of this result, if we choose the exit
direction $\np'$ as the $z$-axis, then the two nucleons at positions
$\nr_1$ and $\nr_2$ are initially at the same height and with the same
impact parameter. Therefore their interaction with the 
central part of the optical potential along its exit trajectory is
the same for both nucleons. 
This proves the persistence of the nodal symmetry in the presence of the
central part of the optical potential.

\item The value of the spin-orbit coupling $\ns\cdot\nl$ for the
ejected particle is the same for both polarizations. 
In fact, the angular momentum of the final proton is
\begin{equation}\label{angular-momentum}
\nl =
\nr\times\np' 
= -\frac{1+\sin^2\theta^*_p}{p^2\sin^2\theta^*_p}
      (\nOmega^*\times\np)\times\np' .
\end{equation}
The double vector product appearing above is given by
\begin{equation}
(\nOmega^*\times\np)\times\np'=
(\nOmega^*\cdot\np')\np-(\np\cdot\np')\nOmega^* .
\end{equation}
The factor in Eq.~(\ref{angular-momentum}) depending on
$\sin\theta^*_p$
is the same for both polarizations.
Therefore the spin-orbit coupling is determined by the following 
scalar product
\begin{equation}
\ns\cdot [(\nOmega^*\times\np)\times\np']=
(\nOmega^*\cdot\np')\np\cdot\ns-(\np\cdot\np')\nOmega^*\cdot\ns .
\end{equation}
Now using Eq.~(\ref{spin}) we have for the spin scalar products:
\begin{eqnarray}
\np\cdot\ns 
& = & \frac{2\nOmega^*\cdot\np}{p^2}\np^2-\nOmega^*\cdot\np 
= \nOmega\cdot\np
\\
\nOmega^*\cdot\ns 
&=&  \frac{2\nOmega^*\cdot\np}{p^2}\np\cdot\nOmega^*-1
=   \left(\frac{2\nOmega^*\cdot\np}{p}\right)^2-1 .
\end{eqnarray}
Therefore 
\begin{equation}
\ns\cdot [(\nOmega^*\times\np)\times\np']=
(\nOmega^*\cdot\np')
(\nOmega^*\cdot\np)
-(\np\cdot\np')
\left[\left(\frac{2\nOmega^*\cdot\np}{p}\right)^2-1\right] .
\end{equation}
Since the projections of $\nOmega_1$ and $\nOmega_2$ on the
scattering plane are opposite, and both $\np$ and $\np'$ are contained
in that plane, we have 
\begin{eqnarray}
(\nOmega_1\cdot\np')
(\nOmega_1\cdot\np)
&=&
(\nOmega_2\cdot\np')
(\nOmega_2\cdot\np)
\\
(\nOmega_1\cdot\np)^2
&=&
(\nOmega_2\cdot\np)^2
\end{eqnarray}
from which it follows that
\begin{equation}
\ns_1\cdot\nl_1=\ns_2\cdot\nl_2 .
\end{equation} 
Hence the spin-orbit interaction is the same for both polarizations.
\end{enumerate}

\section*{Acknowledgments}

This work was partially supported by funds provided by DGICYT (Spain)
under Contract No.  PB/98-1367 and the Junta de Andaluc\'{\i}a
(Spain), and in part by the U.S. Department of Energy under
Cooperative Research Agreement No. DE-FC02-94ER40818.


\end{document}